\documentclass[pdftex,12pt,a4paper]{article}

\usepackage{anysize}
\marginsize{2cm}{2.5cm}{1.5cm}{3.6cm}

\usepackage{relsize}
\usepackage{graphicx}
\usepackage{fancyhdr}
\usepackage{setspace}
\usepackage{amsfonts}
\usepackage{enumerate}
\usepackage{amsthm}
\usepackage{amsmath}
\usepackage{wrapfig}
\usepackage{subfig}
\usepackage{colortbl}

\usepackage{amssymb}
\usepackage{colortbl}

\begin{document}

\newcommand{\ket}[1]{\arrowvert #1 \rangle}
\newcommand{\bra}[1]{\langle #1 \arrowvert}
\newcommand{\scal}[2]{\langle #1 \arrowvert #2 \rangle}
\newcommand{\proj}[2]{\arrowvert #1 \rangle \langle #2 \arrowvert}
\newcommand{\mean}[1]{\langle #1 \rangle}
\newcommand{\sand}[3]{\langle #1 \arrowvert #2 \arrowvert #3 \rangle}
\newcommand{\tr}{\mbox{Tr}}
\newcommand{\vari}{\mbox{Var}}
\newcommand{\spann}{\mbox{Span}}
\newtheorem{theo}{Theorem}
\newtheorem{defin}{Definition}

\title{\titleofpaper}
\begin{center}

\Large \textbf{Paradigms for Quantum Feedback Control}
\\ \vspace{3mm} \normalsize
L. D. T\'oth$^{1,2}$

\vspace{5mm}

\hspace{12mm} $^{1}$ Department of Applied Maths and Theoretical Physics, University of Cambridge, Wilberforce Road, Cambridge CB3 0WA, United Kingdom

\vspace{3mm}

\hspace{12mm} $^{2}$ PRIMALIGHT, Department of Electrical Engineering, King Abdullah University of Science and Technology, Thuwal 23955-6900, Saudi Arabia

\end{center}

\begin{abstract}

In this review paper, we survey the main concepts and some of the recent developments in quantum feedback control. For consistency and clarity, essential ideas and notations in the theory of open quantum systems and quantum stochastic calculus, as well as continuous measurement theory are developed. We give a general description of quantum feedback control, set up a coherent model and compare it to open-loop designs. Objectives which can be achieved by feedback, such as rapid state preparation and purification or entanglement generation are formulated and analyzed, based on the relevant literature. The connection between quantum feedback and quantum chaos is also described and unravelled which, apart from its theoretical curiosity, can shed more light on some of the intrinsic properties of this control paradigm.

\end{abstract}

\newpage
\doublespacing
\tableofcontents

\newpage

\section{Introduction}
\onehalfspacing
In the 21$^{st}$ century we see the advent of true quantum technologies such as quantum computing \cite{quantumcomp} and quantum metrology \cite{quantummetrology} which are expected to outperform any conventional approach. These fields are rapidly evolving and have created the demand for strategies to govern individual quantum systems. Thus, the theory of control of quantum systems has gained tremendous interest recently and is already a huge field. Parallel to the theoretical advances, the experimental side has also seen a blast, especially due to the swift progress in intense and ultrashort laser pulse generation \cite{attosec} which opens up the prospect to observe and manipulate properties of single molecules, solid-state systems or atomic-scale phenomena in real-time. 

There have been numerous paradigms developed for different tasks and control objectives. One of the early ones is open-loop control which relies on the knowledge of the initial quantum system and a well-defined control objective to design control fields without considering feedback from measurements. This can be done coherently, i.e. we use these control fields in a way that does not destroy quantum coherence which was utilized for problems e.g. in quantum chemistry. In coherent control in general, the control operations consist of unitary transformations. However, some quantum systems may not be controllable using only coherent controls. For such uncontrollable quantum systems, it may be possible to enhance the capabilities of quantum control by introducing new control strategies where one is allowed to destroy coherence of the quantum systems during the control process (incoherent control). Optimal control techniques such as gradient-free convex optimization can also boost the convergence properties and efficiency of open-loop control designs.

Although open-loop strategies have achieved theoretically and practically significant success, they are quite limited in scope. It was natural to extend the studies to closed-loop control which has been investigated in depth in classical control theory and shown to be superior in many ways, most notably in reliability and robustness. These are essential in quantum control because any practical quantum technology - a quantum computer, for instance - has to be robust in the presence of noise or uncertainty. In closed-loop control, the state information is used in shaping the control mechanism. We may split this paradigm into two categories (although other categorizations also exist): adaptive learning control and quantum feedback control (QFC). In the former case we have a closed-loop operation and each cycle is applied on a new sample. This procedure has gained great success where multiple samples are available, e.g. controlling molecules in an ensemble with lasers \cite{moleculecontrol1, moleculecontrol2}.

The other concept, quantum feedback control (QFC) includes direct or indirect measurements on the state to gain information which can be fed back to achieve the desired performance. Classical feedback control is well-understood and has tremendous advantages because - in principle - the measurement back-action can be neglected classically, i.e. we can acquire full information without disturbing the system. Due to the intrinsically different nature of quantum mechanics, most importantly the well-known phenomena of quantum state collapse and the unnegligable measurement back-action, QFC faces a great number of challenges. Nevertheless, much has been done since the first recognition (see e.g. \cite{belavkin}) of the importance of this paradigm and promising results have been obtained.

The paper is organized as follows. In Section \ref{sec:formalism}., we overview the most important elements of the general framework needed to formulate problems in quantum feedback control such as the Markovian master equation, weak and Continuous measurements, stochastic Schr\"odinger and master equations \cite{opensystems,weakmeasurement2,weakmeasurement1}. This can be useful for people who are not familiar with the formalism of this field and also helps the survey to be self-consistent, self-contained and to avoid ambiguities in notation. In Section \ref{sec:gendesc}., we give a general description of QFC and in Section \ref{sec:cohmodell}., we set up a coherent control model \cite{modellingfeedbackcontrol} and outline some theorems regarding its capability \cite{measurementbasedfeedback1}. Furthermore, in a generalization of this setup, it is convenient to compare the results with open-loop control \cite{reachabilityofmfc}. In Section \ref{sec:quantumchaos}., we briefly introduce how quantum feedback can help us to understand quantum chaos better which has found to be essential in understanding the transition from quantum to classical \cite{chaosbook,habibetal,slotine,complexqubit1}. In the following section we review some of the tasks which are proven to be efficiently achievable such as quantum error correction, rapid state preparation and purification, entanglement generation. We start with a simple example: feedback control of a single qubit in a discrete-time setting \cite{rapidprepqubit}. This clearly illustrates the central concepts. Then we move on to a similar task but with a Continuous measurements \cite{rapidprepqubit}; at the end of this section, we also make some remarks on recently emerged questions (arbitrary large systems \cite{rapidstatereduction}, feedback delay problem \cite{openloopfiltering}). Entanglement generation is a novel example where quantum feedback is useful \cite{entgen1,entgen2,entgen3} which is described in Section \ref{sec:entgen}. A projective measurement-based feedback scheme is described in Section \ref{sec:projqubit} which connects chaos and quantum feedback control from a different perspective. This scheme can be used for several purposes such as state purification or enhancing entanglement (which has been proven for a two-qubit case \ref{sajat}). We present reproduced simulations in some of the cases (Section \ref{sec:projqubit}).

\section{Background and formalism}\label{sec:formalism}
\subsection{Quantum mechanics, strong and weak measurements}\label{subsec:measurements}

In quantum control the systems we want to control are quantum systems, thus described by the general framework of quantum mechanics. For closed systems the state is described by a unit vector $\ket{\psi}$ in a Hilbert space $\mathcal{H}$ (\bfseries state space postulate\mdseries). The time-evolution of the state of a closed quantum system is described by a unitary operator (\bfseries evolution postulate\mdseries). When two physical systems are treated as one combined system, the state space of the combined physical system is the tensor product space $\mathcal{H}_1 \otimes \mathcal{H}_2$ of the state spaces $\mathcal{H}_1, \mathcal{H}_2$ of the component subsystems (\bfseries composition of systems postulate\mdseries). Quantum measurements are described by a set of measurement operators $\{M_m\}$ which act on the state space of the system being measured and satisfy $\sum_m M_m^{\dag} M_m=I$ (the index $m$ refers to the measurement outcomes). If the state of the system immediately before the measurement is $\ket{\psi}$ then the probability that we measure $m$ is $\bra{\psi}M_m^{\dag} M_m \ket{\psi}$ and the measurement leaves the system in $(\bra{\psi}M_m^{\dag} M_m \ket{\psi})^{-1/2} \cdot M_m \ket{\psi}$ (\bfseries measurement postulate\mdseries).

In many situations we only know a probability distribution about the states in which the system can be, i.e. $\{(\ket{\psi_k}, p_k)\}_{k=1,2,...} \;\;\;\; 0 \leq p_k \leq 1 \;\;\;\; \sum_k {p_k}=1$ (the system is in a \itshape  mixed state\upshape ). In this case, it is convenient to introduce the density operator: $\hat{\rho} \equiv \sum_k{p_k \ket{\psi_k}\bra{\psi_k}}$. From construction we know that $\hat{\rho}$ is Hermitian, positive and $\tr(\hat{\rho})=1$. Also, $\tr(\hat{\rho}^2)=1$ if and only if the system is in a pure state and under a $U$ unitary transformation $\hat{\rho}$ transforms as $\hat{\rho} \longrightarrow \hat{\rho} ' = U \hat{\rho} U^{\dag}$.

Let us introduce now another class of measurements which provide only partial information about an observable. We can do this if we choose our measurement operators to be a weighted sum of projectors (we will denote their eigenstates as $\ket{n}$), each on peaked about a different value of the observable, i.e.
\begin{equation} \label{eq:weightedsummeasurement}
M_m = \dfrac{1}{\mathcal{N}} \sum_n e^{-k(n-m)^2/4} \ket{n} \bra{n}
\end{equation}
where $\mathcal{N}$ is the normalization factor chosen so that $M_m$ satisfy the completeness relation $\sum_m M_m^{\dag} M_m=I$ and we assumed that the eigenvalues of the observable N are $n \in \mathbb{Z}$. As an example, if apply this measurement to a completely mixed state (so $\rho \propto I$) and obtain the result $m$, the post-measurement state is
\begin{equation} \label{eq:postmeasurementstate}
\rho' = \dfrac{1}{\mathcal{N}} \sum_n e^{-k(n-m)^2/2} \ket{n} \bra{n}
\end{equation}
from which we see that the final state is peaked about the eigenvalue $m$ but has a finite width given by $1/\sqrt{k}$. Measurements for which $k$ is large are called \itshape strong measurements \upshape and for which $k$ is small are called \itshape weak measurements\upshape \cite{weakmeasurement1,weakmeasurement2}.

\subsection{Open quantum systems}\label{subsec:openquantumsystems}

The above postulates might be enough to treat closed quantum systems; however, in most practical situations we have to deal with open systems (for a deep introduction we refer to \cite{opensystems}). On one hand, this is due to the fact that any realistic system is subjected to a coupling to a second system	(we will use the term \itshape environment \upshape or \itshape bath \upshape for the second system if it is much larger than the first) in an uncontrollable and non-negligable way. On the other hand, even if one can provide and solve a microscopic description for the combined system, most of the results would be irrelevant. Open systems are also important when we want to monitor (i.e. Continuously measure) a system. Both cases are essential in quantum feedback control.

We start with a closed system. The Schr\"odinger equation is
\begin{equation} \label{eq:schrodinger}
i \dfrac{d}{dt}\ket{\psi(t)} = \hat{H}(t)\ket{\psi(t)}
\end{equation}
where $H(t)$ is the Hamiltonian of the system and we have set $\hbar$ to $1$. It is easy to see that for a mixed state (\ref{eq:schrodinger}) implies
\begin{equation} \label{eq:neumanneq}
\dfrac{d}{dt}\hat{\rho}(t) = -i[\hat{H}(t), \hat{\rho}(t)]
\end{equation}
which is called the von Neumann or Liouville - von Neumann equation. This can be written in the form
\begin{equation} \label{eq:neumannliouville}
\dfrac{d}{dt}\hat{\rho}(t) = \hat{\mathcal{L}}(t)\hat{\rho}(t)
\end{equation}
where one can easily notice the analogy with the classical Liouville equation. Here, $\hat{\mathcal{L}}$ is the Liouville super-operator (it is a super-operator, since it maps operators to operators). We will drop the hats from the operators from now on. Note, that if we work in the interaction picture, (\ref{eq:neumanneq}) and (\ref{eq:neumannliouville}) still hold for the interaction density matrix and interaction Hamiltonian (which we will denote with a subscript $I$). We can write (\ref{eq:neumanneq}) in an integral form as
\begin{equation} \label{eq:formalsolutionforneumanneq}
\rho_I(t)=\rho_I(t_0) - i \int_{t_0}^{t}{[H_I(t'), \rho(t')] dt'}
\end{equation}
We can use this to solve (\ref{eq:neumanneq}) perturbatively.

Formally, it is easy to generalize (\ref{eq:neumanneq}) to open systems. Consider the case when the system is in a bath (thus, the Hilbert space of the total system is $\mathcal{H}_{total}=\mathcal{H}_{system} \otimes \mathcal{H}_{bath}$). The Hamiltonian of the total system can be written $H(t)_{total}=H_{system} \otimes I_{bath} + I_{system}\otimes H_{bath} + H_{interaction} (t)$. The density matrix of the system can be obtained by tracing out the bath from the density matrix describing the total system (S+B): $\rho_{S}=Tr_{B}(\rho_{S+B})$. So (\ref{eq:neumanneq}) takes the form
\begin{equation} \label{eq:neumanneqforopensystems}
\dfrac{d}{dt}\rho_S(t) = -i \tr_B [H(t), \rho(t)]
\end{equation}
However, in general the dynamics of $\rho_S$ can be rather involved and we have to make assumptions to proceed. Let us assume that at $t=0$ the system is uncoupled from the environment, i.e. $\rho(0)= \rho_S(0) \otimes \rho_B$ where $\rho_B$ represents some reference state of the bath. Then the transformation from $t=0$ to some $t>0$ of $\rho_S$ can be written as $\rho_S(0) \rightarrow \rho_S(t)=V(t)\rho_S(0)$. We introduced $V(t)$ which is a map from the system space to itself and is called a  \itshape dynamical map\upshape. It can be shown that these maps represent convex-linear, completely positive and trace-preserving quantum operations. If we neglect the memory effects in the reduced system dynamics (justified later) we can show that they also form a semigroup. Under some mathematical conditions \cite{opensystems} there exists a linear map (let us call it $\mathcal{L}$) which is the generator of the semigroup, so we can write $V(t)=exp(\mathcal{L}t)$. From this we rewrite (\ref{eq:neumannliouville}) as
\begin{equation} \label{eq:markovianqme}
\dfrac{d}{dt}\rho_S(t) = \mathcal{L} \rho_S(t)
\end{equation}
which is called the \itshape Markovian master equation\upshape. It was shown by Lindblad in 1976 that the most general form of $\mathcal{L}$ (so that a solution is always a valid density matrix) is (assuming that the dimension of the Hilbert space of the total system is $N<\infty$)
\begin{equation} \label{eq:generalgenerator}
\mathcal{L} \rho_S = -i [H, \rho_S] + \sum_{k=1}^{N^2 - 1} \gamma_k \left(L_k \rho_S L_k^{\dag} - \dfrac{1}{2} L_k^{\dag}L_k \rho_S - \dfrac{1}{2}\rho_S L_k^{\dag}L_k \right)
\end{equation}
where the quantities $\gamma_k$ are non-negative (and can be shown that physically they play the role of relaxation rate for the different decay modes of the system), the operators $L_k$ are arbitrary operators, called the \itshape Lindblad operators\upshape, satisfying that $\sum_k L_k^{\dag} L_k$ is bounded (although this condition is usually ignored). The first term represents the unitary part of the dynamics and the second term is the dissipative part (often denoted as $\mathcal{D}[\rho_S]$).

It is possible to derive the generator $\mathcal{L}$ assuming various underlying Hamiltonian dynamics, although several approximations are usually needed. As an example, if we consider a weakly coupled system and use the Born-Markov approximation \cite{bornmarkov}, we obtain the \itshape Born-Markov quantum master equation\upshape
\begin{equation} \label{eq:bornmarkov}
\dfrac{d}{dt}\rho_S(t) = -\int_0^{\infty} \tr_B[H_I(t), [H_I(t-t'), \rho_S(t) \otimes \rho_B]] dt'
\end{equation}
It is worth mentioning what physical assumptions we made here. First, we assumed that the bath is only negligibly affected by the interaction. Secondly, we said that environmental excitations decay over times which are not resolved (in other words, the time scale over which the state of the system varies appreciably is large compared to the time scale over which the reservoir correlation functions decay).

\subsection{Continuous measurements and stochastic processes}\label{sec:stoch}

Here we use the formalism introduced in Section \ref{subsec:measurements}. to describe Continuous measurements and to derive the stochastic Schr\"dinger equation and the stochastic master equation. The reader is referred to \cite{weakmeasurement2} for a broader introduction but we closely follow its lines. These are the most commonly used equations in quantum feedback control theory.

Continuous measurement means that we continually extract information from the system and it can be obtained by performing a weak measurement in $\Delta t$ time steps with the measurement strength being also proportional to $\Delta t$, then taking the limit $\Delta t \rightarrow 0$. Let $X$ be a Hermitian operator (i.e. an observable) and for simplicity assume that $X$ has a Continuous spectrum (it turns out that the results are valid for any Hermitian operator\cite{weakmeasurement2}). Denote the eigenstates as $\ket{x}$, so $\scal{x}{x'}=\delta(x-x')$. Now, in analogy with (\ref{eq:weightedsummeasurement}), let us perform a measurement described by the operator
\begin{equation} \label{eq:weightedsumContinuous}
M(\mu)=\left(\dfrac{4k\Delta t}{\pi}\right)^{1/4} \int_{-\infty}^{\infty} e^{-2k \Delta t (x - \mu)^2} \proj{x}{x} dx
\end{equation}
at every time step $\Delta t$, where $M(\mu)$ is a Gaussian-weighted sum of projectors onto the eigenstates of $X$. Choosing the initial state as $\ket{\psi}=\int{\psi(x) \ket{x} dx}$ the probability density $P(\mu)$ of the measurement result $\mu$ is
\begin{equation} \label{eq:probdensofmu}
P(\mu)=\tr[M(\mu)^{\dag} \proj{\psi}{\psi} M(\mu)]=\left(\dfrac{4k\Delta t}{\pi}\right)^{1/2} \int_{-\infty}^{\infty} e^{-4k \Delta t (x - \mu)^2} |\psi(x)|^2 dx
\end{equation}
and the mean value of $\mu$ is
\begin{equation} \label{eq:meanofmu}
\langle \mu \rangle = \int_{-\infty}^{\infty} \mu P(\mu) d\mu  = \int_{-\infty}^{\infty} x|\psi(x)|^2 dx = \langle X \rangle
\end{equation}
where we simply plugged in (\ref{eq:probdensofmu}) and used $\int_{-\infty}^{\infty} \mu e^{-4k \Delta t (x - \mu)^2} d \mu=(4 k \Delta t)^{-1/2}(\pi)^{1/2}\cdot x$ from where we can also justify our chosen normalization. Now we approximate (\ref{eq:probdensofmu}): we replace $|\psi(x)|^2$ by a $\delta$-function centered at $\langle X \rangle=\langle \mu \rangle$. This can be justified because if $\Delta t$ is sufficiently small, the Gaussian is much broader than $\psi(x)$. After the substitution and performing the trivial integral we obtain
\begin{equation} \label{eq:probdensofmuapprox}
P(\mu) \approx \left(\dfrac{4k\Delta t}{\pi}\right)^{1/2} e^{-4k \Delta t (\mu - \mean{X})^2}
\end{equation}
We can represent $\mu$ as a stochastic variable
\begin{equation} \label{eq:mustoch}
\mu_s = \mean{X} + \dfrac{\Delta W}{\sqrt{8k\Delta t}}
\end{equation}
where $\mean{\Delta W}=0$ and its variance is $\vari(\Delta W)= \Delta t$. $\Delta W$ is called the Wiener increment and it is worth stopping here and defining it in a more general fashion. The Wiener process $W(t)$ is a Continuous-time stochastic process with three properties: $W(0)=0$, $W(t)$ is almost surely Continuous and $W(t)$ has independent increments with $W(t)-W(s) \propto \mathcal{N}(0, t-s) \;\;\; \forall 0 \leq s < t$ where $\mathcal{N}(m, \sigma^2)$ denotes the normal distribution with expected value $m$ and variance $\sigma^2$. The condition that it has independent increments means that if $0 \leq s_1 < t_1$ and $0 \leq s_2 < t_2$ then $W(t_1)-W(s_1)$ and $W(t_2)-W(s_2)$ are independent random variables and similar condition holds for $n$ increments. It can be shown \cite{stochcalc} that at a fixed time $t$, the probability density function of $W(t)$ is
\begin{equation} \label{eq:probdensfuncw}
P[W(t)]=\dfrac{1}{\sqrt{2 \pi t}}e^{-x^2/2t}
\end{equation}
and also the expectation value of $W(t)$ is zero and the variance is $t$. An important rule for the Wiener increment is that when we take the infinitesimal limit (i.e. $\Delta t \rightarrow dt$ and $\Delta W \rightarrow dW$) then 
\begin{equation} \label{eq:itowiener}
(dW)^2=dt
\end{equation}
This is one of the so called $It\hat{o} \; rules$. The outline of the proof of this statement is as follows. We consider the probability  density function of $(\Delta W)^2$ which we obtain by a simple transformation of (\ref{eq:probdensfuncw}). From this we can see that $\mean{(\Delta W)^2}=\Delta t$ and $\vari((\Delta W)^2)=2(\Delta t)^2$. We split up $\Delta t$ into $N$ intervals and sum of the squares of all the corresponding Wiener increments, i.e.  $(\Delta W_n)^2$. Now if we take the continuum limit, this sum becomes a Gaussian random variable (from the central limit theorem) with mean $t$ and variance $2t^2/N$. As $N \rightarrow \infty$, this variance vanishes. So
\begin{equation} \label{eq:itowiener2}
\int_0^t (dW(t'))^2 \stackrel{!}{=} \lim_{N \rightarrow \infty} \sum_{n=0}^{N-1} (\Delta W_n)^2=t=\int_0^t dt'
\end{equation}
We want this to hold for any interval $(0,t)$, so we must have $dt=dW^2$. This also means that $dW^2$ is not a random variable since it has no variance when integrated over any finite inverval.

Now let us examine the equation of motion conditioned upon our measurement (\ref{eq:weightedsumContinuous}). The evolution can be computed (using only the postulates) as
\begin{equation} \label{eq:condevo1}
\ket{\psi(t + \Delta t)} \propto M(\mu) \ket{\psi(t)} \propto e^{-2k\Delta t (\mu_s - X)^2}\ket{\psi (t)} \propto e^{-2k\Delta t X^2 + X(4k \mean{X}\Delta t + \sqrt{2k}\Delta W)}\ket{\psi (t)} 
\end{equation}
where we simply applied the measurement $M(\mu)$ at a time step $\Delta t$ and ignored the normalization and other constant factors. Now expand (\ref{eq:condevo1}) to first order in $\Delta t$ (which means that we have to keep second order terms in $\Delta W$ according to the Ito-rule):
\begin{equation} \label{eq:condevo2}
\ket{\psi(t + \Delta t)} \propto [1 - 2k\Delta t X^2 + X(4k \mean{X} \Delta t + \sqrt{2k}\Delta W + kX(\Delta W)^2)] \ket{\psi (t)} 
\end{equation}
Now we take the continuum limit, so $\Delta t \rightarrow dt \;\;\; \Delta W \rightarrow dW \;\;\; (\Delta W)^2\rightarrow dt$ and we have
\begin{equation} \label{eq:condevo3}
\ket{\psi(t + \Delta t)} \propto [1 - (kX^2 - 4kX \mean{X})dt + \sqrt{2k}X dW] \ket{\psi (t)} 
\end{equation}
However, we must be aware of the fact that - for simplicity - we derived this without considering the normalization (thus we wrote $\propto$). It is straightforward to take into account the normalization and we obtain
\begin{equation} \label{eq:stochasticschrodinger}
d\ket{\psi}= [-k(X - \mean{X})^2 dt + \sqrt{2k}(X - \mean{X}) dW] \ket{\psi (t)} 
\end{equation}
where $d\ket{\psi}=\ket{\psi(t + \Delta t)}-\ket{\psi(t)}$. This is the well-known \itshape stochastic Schr\"odinger equation \upshape (SSE), although not in its most general form (e.g. we chose the measurement operator to be Hermitian). In terms of the density operator this gives
\begin{equation} \label{eq:stochasticschrodingerdensity}
d\rho = -k[X,[X, \rho]]dt + \sqrt{2k}(X\rho + \rho X - 2\mean{X}\rho)dW
\end{equation}
which is the \itshape stochastic master equation \upshape (SME) \cite{belavkin} without the Hamiltonian evolution term. The first term in the equation describes the drift towards the measurement axis. The second term describes the update of knowledge of the density matrix conditioned on the measurement. The measurement record is explicitly
\begin{equation} \label{eq:meausrementrecord}
dy = \mean{X} dt + \dfrac{dW}{\sqrt{8k}}
\end{equation}
in a time interval $dt$. $\ket{\psi(t)}$ and $\rho(t)$ define a quantum trajectory. We can write (\ref{eq:stochasticschrodingerdensity}) in terms of the the measurement record (\ref{eq:meausrementrecord}) and we have
\begin{equation} \label{eq:smem}
d\rho = -k[X,[X, \rho]]dt + 4k(\{X, \rho \} -2\mean{X} \rho)(dy - \mean{X}dt)
\end{equation}
where the $\{\}$ brackets denote the anticommutator. It is worth mentioning that if one performs the Continuous measurement but throws away the record then the SME becomes
\begin{equation} \label{eq:smenomeasure}
\dfrac{d\rho}{dt} = -k[X,[X, \rho]]
\end{equation}
which can be obtained by averaging over all the possible records and noting that $\rho$ and $dW$ are statistically independent (so we can throw away the $dW$ term in the SME).

To have the whole picture, we extend the SME to a more general form, without derivation - it would be fairly straightforward from the Markovian master equation (\ref{eq:markovianqme}) - but interpreting the terms. The most general form of the SME with Wiener noise and without taking into account that the noise sources can also be complex and mutually correlated is
\begin{equation} \label{eq:generalsme}
d\rho = -i[H, \rho]dt + \sum_n \mathcal{D}[c_n] \rho dt + \sqrt{\eta_n}\mathcal{H}[c_n] \rho dW
\end{equation}
where $c$ is an arbitrary operator, $\mathcal{D}[c]\rho=c\rho c^{\dagger}-\frac12(c^{\dagger}c\rho + \rho c^{\dagger}c)$ is the dissipation superoperator, $\mathcal{H}[c]\rho=c\rho + \rho c^{\dagger} - \mean{c + c^{\dagger}}\rho$ is the measurement superoperator, we allowed $n$ number of measurements (\itshape output channels\upshape) and $\eta_n$ is the efficiency of the $n$th detection channel \cite{weakmeasurement2,weakmeasurement3}.

Let us also introduce the Ito-integral and its transformation properties. We often encounter the following stochastic differential equation:
\begin{equation} \label{eq:itoeq}
\dfrac{d X_t}{dt}=f(X_t) + \sigma (X_t)\dfrac{dW_t}{dt}
\end{equation}
$W_t$ is a one-dimensional Wiener process. This, however, makes little sense as $W_t$ is not differentiable. We can resolve this problem if we write (\ref{eq:itoeq}) in an integral form
\begin{equation} \label{eq:itoeqintform}
X_t = X_0 + \int_0^t f(X_s) ds + \int_0^t \sigma(X_s) dW_s
\end{equation}
or, in analogy with ordinary differential equations
\begin{equation} \label{eq:itoeqintformanalogy}
dX_t = f(X_t)dt + \sigma(X_t) dW_t
\end{equation}
and then define the second integral in (\ref{eq:itoeqintform}). There are several possible ways to do it (e.g. the Stratonovich form, Ito form), we consider the latter:
\begin{equation} \label{eq:itoint}
\int_{t_0}^{t_n} f_s dW_s = \lim_{|t_{i+1} - t_i| \rightarrow 0} \sum_{k=0}^{n-1}f_{t_k}(W_{t_{k+1}}-W_{t_k})
\end{equation}
A very important remark is that the Ito integral does not obey the usual Leibniz rule which is $d(X_t Y_t)=X_t dY_t + Y_tdX_t$, instead it obeys $d(X_t Y_t)=X_t dY_t + Y_tdX_t + dX_t dY_t$. Similarly, for an arbitrary function of $X_t$ we have $g(X_t)=g'(X_t)dX_t + \frac12 g''(X_t)dX_t^2$.

\newpage

\section{Quantum feedback control}\label{sec:qfc}
\subsection{A general description}\label{sec:gendesc}

As it was mentioned before, feedback control is an ubiquitous and powerful technique for classical systems because - in principle - it is possible to acquire all the information about the state of a system with certainty by using sufficiently precise measurements. However, there are two fundamental features of quantum systems which have to be taken into account in the quantum case. The first one is that non-orthogonal states cannot be distinguished with certainty. The second one is that any measurement that gains information about a system induces an uncontrollable noise to it. Therefore, one must carefully design the control scheme to balance the trade-off between information gain and disturbance.

\begin{wrapfigure}{r}{0.5\textwidth}
\vspace{-20pt}

  \begin{center}
    \includegraphics[width=0.48\textwidth]{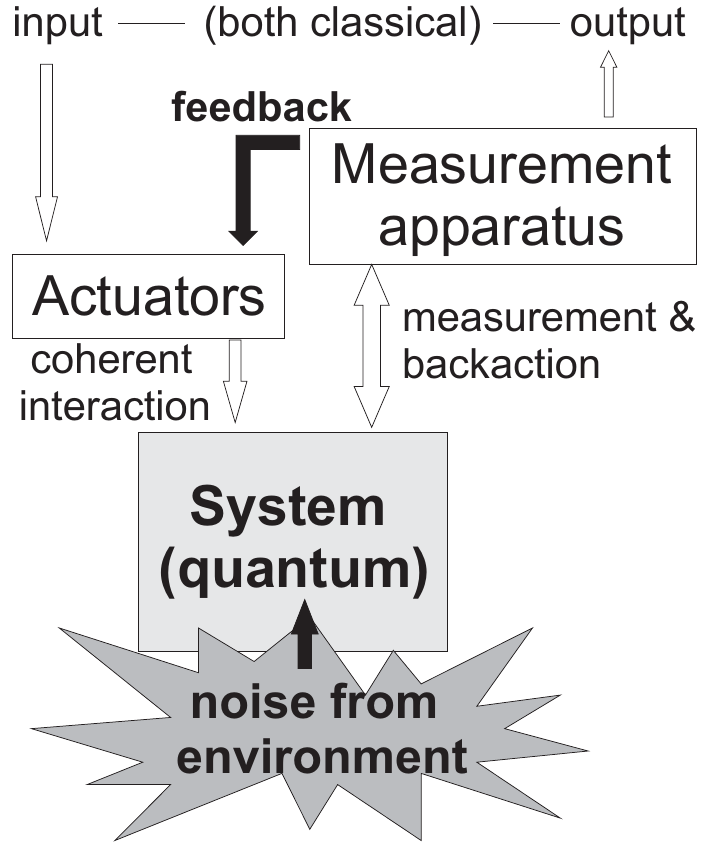}
  \end{center}
  \vspace{-20pt}
  \caption{A schematic diagram of quantum feedback control, reproduced from \cite{quantumcontrolcourse}.}
  \vspace{-10pt}
\end{wrapfigure}

QFC can be split up into several subcategories depending on the type of measurement we use or the way we treat the controller. The two main approaches to information acquistion are strong (projective) measurements and Continuous weak measurements. The controller can be considered as a classical object - that is, the gained information is classical - or quantum system which processes -and feeds back - quantum information \cite{coherentquantumfeedback}. If we assume, that the controller is memoryless and we do not consider any time delays (so e.g. we can immediately feed back the information) then it is Markovian feedback control and the resulting evolution of the system is described by the Markovian master equation (\ref{eq:markovianqme}) derived in Section \ref{subsec:openquantumsystems}. In contrast, we can devide the whole control process into two steps: first, we estimate some of the dynamical variables of the state and use the estimated state to design the control. It is usually desirable to obtain the measurement record Continuously, therefore this technique requires real-time solution of stochastic differential equations and fast measurements. The dynamical equation of the evolution is non-Markovian. Determining the conditioned state of the quantum system from classical measurement results is a quantum version of Bayesian reasoning. Classical Bayesian reasoning updates an observers knowledge of a system (as described by a probability distribution over its variables) based on new data. For this reason, this feedback paradigm is called Bayesian feedback. In \cite{bayesian} one can find a good comparison of the Markovian and Bayesian feedback; the results prove that the latter is never inferior, and is usually superior, to the former. However, it would be far more difficult to implement than Markovian feedback and it loses its superiority when obvious simplifying approximations are made.

An interesting, somewhat hybrid approach is the Lyapunov-method which first constructs an artificial closed-loop controller, simulates it (e.g. on a computer) and the open-loop control law is obtained by the result ("feedback design and open-loop control strategy"). The most important aspects of this methodology are: construction of the Lyapunov-function, the determination of the control law and the analysis of the asymptotic convergence. For example, if our desired final state is $\ket{\psi}_t$, we can choose $V(t)=\frac{1}{2}(1-|\bra{\psi_t}\ket{\psi}|^2)$ as a Lyapunov-function and construct the control law to guarantee that $\frac{d}{dt}V(t) \leq 0$. In this paper, we do not consider this method further.

\subsection{An all-optical, coherent model and comparison to open-loop control}\label{sec:cohmodell}

In quantum optics it is a common scenario that the system to be controlled is brought in weak interaction with an external probe field which is subsequently detected \cite{quantummeasurementandcontrol}. Here we consider a system which is an ensemble of atomic spins interacting dispersively with an optical probe which is subjected to homodyne detection. This is an illustrative set-up in several aspects and was experimentally implemented \cite{spinsqueezexp} and theoretically analized for feedback control\cite{modellingfeedbackcontrol,feedbackforstateprepmat}. As the detailed derivation from first principles would abundantly exceed the limit of this paper, the stress will be on describing the physical model and key assumptions, rather than the technicalities needed, to obtain the the SME and other main results, which we can further analize and compare it to open-loop control.

We put the atomic spin ensemble in a leaky single-mode cavity and take the problem to be one-dimensional (which can be justified if we consider that most of the light is scattered forward). The strong driving field is treated semiclassically and polarized light is assumed. The leaky single-mode cavity description also allows us to treat the interaction between the ensemble and the field with a single frequency, which is chosen to be the laser frequency $\omega_0$. Spontanious emission into the eliminated modes can be added to the model phenomenologically. In this model the total Hamiltonian can be written as
\begin{equation} \label{eq:totalhamiltonian}
H= H_A + H_D + H_{CF} + H_{AC} + H_{SE}
\end{equation}
where $H_A$ is the atomic Hamiltonian (which can be quite involved, depending on the structure of the atoms in the ensemble), $H_D$ is the interaction Hamiltonian of the cavity mode, $H_{CF}$ is the interaction Hamiltonian between the cavity mode and the external field, given by
\begin{equation} \label{eq:cfhamiltonian}
H_{CF}=\int_0^{\infty} \kappa(\omega)\left(i a^{\dagger}(\omega)be^{i(\omega - \omega_0)t} + i a^{\dagger}(\omega)b^{\dagger}e^{i(\omega + \omega_0)t} + h.c.\right) d\omega
\end{equation}
where $a(\omega),a^{\dagger}(\omega)$ are the annihilators/creators for the electric field (they correspond to plane wave modes in the $z$-direction) and $b(t),b^{\dagger}(t)$ are the cavity mode annihilators/creators with $b(t)=be^{-i\omega_0 t}$, $\kappa(\omega)$ is the mode function. The remaining Hamiltonians are: $H_{AC}$ is the ensemble-cavity mode interaction Hamiltonian and $H_{SE}$ corresponds to spontanious emission. Note that the quadratures $a(\omega) + a(\omega)^{\dagger}$ and $ia(\omega) - ia(\omega)^{\dagger}$ are Gaussian random variables. We will call
\begin{equation} \label{eq:atime}
a(t)=\dfrac{1}{\sqrt{2\pi}}\int_{-\infty}^{+\infty} a(\omega) e^{-i\omega t} d\omega
\end{equation}
quantum white noise as - in the vacuum state - its quadratures have zero mean and delta-correlated covariance. Also note, however, that the whole evolution described by (\ref{eq:totalhamiltonian}) is not driven by white noise (i.e. with a fast varying $\kappa(\omega)$). We have to make a key assumption to obtain this, namely that the cavity is weakly coupled to the external field. We can define
\begin{equation} \label{eq:atimewiener}
A_t=\int_0^{\infty}a(t) dt
\end{equation}
which is a Wiener process and we can define integrals $\int X_s dA_s$ as they were defined in (\ref{eq:itoint}).

Now we introduce a model for the atomic ensemble. Consider $N$ atoms with a degenerate two-level ground state. Assume also that all the atomic transitions are far detuned from the cavity resonance; $F_z$ will denote the collective dipole moment (so it is a spin-N/2 angular momentum operator). Thus, $H_{AC}= \chi F_z b^{\dagger}b$ where $\chi$ is the coupling strength and $H_A=\Delta F_z + u(t)F_y$ where $\Delta$ is the atomic detuning and $u(t)$ is the strength of the magnetic field in the $y$-direction (which is under our control). If we do not consider feedback control, the open-loop evolution can be written as
\begin{equation} \label{eq:olcevolutioncoh}
\dfrac{\rho (t)}{dt}= -i[\Delta F_z + u(t)F_y, \rho(t)]
\end{equation}
We can add the decoherence term due to spontanious emission in a phenomenological way as follows. Introduce another field $\tilde{E}$ with annihilators $c(\omega)$ called the side channel. This will be left unobserved. So $H_{SE}=-d(t)\tilde{E}(0,t)$ where $d(t)$ is the atomic dipole operator. Because the atoms are coupled directly to $\tilde{E}(0,t)$, we can write $d(t)=\sigma e^{-i\omega_d t} + \sigma^{\dagger} e^{i\omega_d t}$ where $\sigma$ is the atomic decay operator and $\omega_d$ is the dipole rotation frequency. This modifies (\ref{eq:olcevolutioncoh}) with an extra term:
\begin{equation} \label{eq:olcevolutiondecoh}
\dfrac{\rho (t)}{dt}= -i[\Delta F_z + u(t)F_y, \rho(t)] + \gamma \mathcal{D}[\sigma]\rho(t)
\end{equation}
where $\gamma$ is the decoherence strength and $\mathcal{D}[\sigma]\rho=\sigma \rho \sigma^{\dagger} -\frac12( \sigma^{\dagger} \sigma \rho + \rho  \sigma^{\dagger}  \sigma)$ is the dissipative superoperator, just as defined in (\ref{eq:generalsme}).

\begin{figure}[h!]
\begin{center}
\includegraphics[width=17cm]{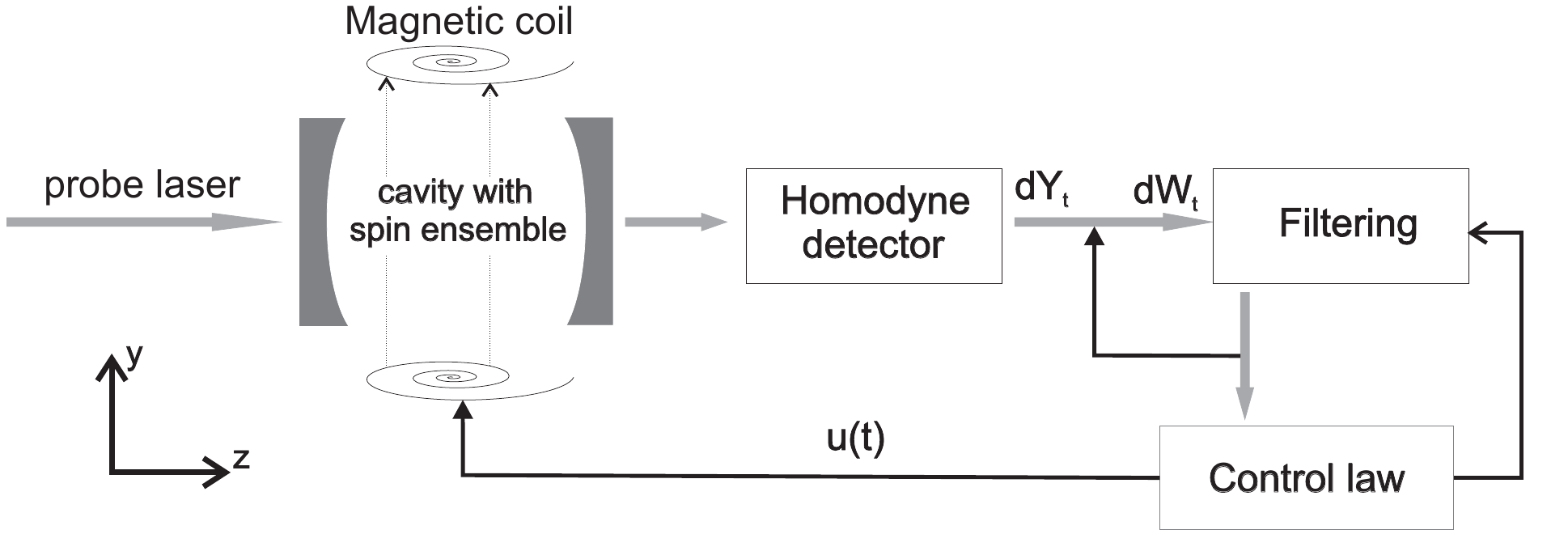}
\caption{The feedback setup for the spin ensemble. The system is in a leaky single-mode cavity and it interacts with a laser field. After the interaction the beam is measured by a homodyne detector and processed. This drives the filtering process which produces the best mean square estimate of the system state. This is used directly to design the control law. The feedback is achived by applying magnetic field.}
\label{fig:feedbackproc}
\end{center}
\end{figure}

Now we introduce our measurement and feedback procedure. Homodyne detection is a powerful technique in practical detection of light beams. This involves another strong, coherent signal (called the local oscillator which, in the homodyne case, has the same frequency as that of the detected signal) which is mixed with the original signal. This can help to eliminate the initial fluctuations of the laser and allows us to detect a quadrature of the system, e.g. $a_t + a^{\dagger}_t$ after the field has interacted with the spins; thus, we observe the photocurrent $I(t)=U_t^{\dagger}(a_t + a^{\dagger}_t)U_t$ where $U_t$ is the evolution of the whole system. It is useful to introduce the integral form of $I(t)$, the integrated photocurrent $Y_t$ which is then $Y_t=U_t^{\dagger}(A_t + A^{\dagger}_t)U_t$. Using this we have to solve the quantum filtering problem. That is, given an atomic observable $X$, we want to find the best estimate of $X$ given the prior observations, formally $\mathbb{E}[U_t^{\dagger} X U_t | Y_{s \leq t}]$. This problem can be solved after deriving the whole evolution of the system and by using properties of the conditional expectation, tricks form real analysis and the Ito rules. In this model the SME derived in (\ref{eq:generalsme}) takes the form \cite{modellingfeedbackcontrol}:
\begin{equation} \label{eq:mfcevolutionwodec}
d\rho_t=-iu(t)[F_y, \rho_t]dt - is[F_z, \rho_t]dt + M \mathcal{D}[F_z]\rho_t dt + \sqrt{M\eta}\mathcal{H}[F_z]\rho_t dW_t
\end{equation}
where $s$ is a parameter which depends on the experimental setup (such as $\chi$), $\eta\in(0,1]$ is the detection efficiency, $M$ is the effective interaction strength which is a function of $\chi$ and the drive amplitude (this quantity manifests itself in the Hamiltonian of the interaction of the cavity) and, as a reminder, $\mathcal{H}[F_z]\rho=\rho=F_z\rho + \rho F_z^{\dagger} - \mean{F_z + F_z^{\dagger}}\rho$ and we defined $dW_t$ (innovation process) as
\begin{equation} \label{eq:mfcinnovproc}
dW_t = dY_t - 2\sqrt{M\eta} \mean{F_z} dt
\end{equation}
and it can be shown that it is in fact a Wiener process (the idea is the same as in Section \ref{sec:stoch}.). It contains white noise terms and also the new information provided by the measurement. Just as in the open-loop case, we can add the decoherence term describing the spontanious emission and we have
\begin{equation} \label{eq:mfcevolutionwithdec}
d\rho_t=-iu(t)[F_y, \rho_t]dt - is[F_z, \rho_t]dt + M \mathcal{D}[F_z]\rho_t dt + \gamma \mathcal{D}[\sigma]\rho_t + \sqrt{M\eta}\mathcal{H}[F_z]\rho_t dW_t
\end{equation}
With equations (\ref{eq:olcevolutioncoh}), (\ref{eq:olcevolutiondecoh}), (\ref{eq:mfcevolutionwodec}) and (\ref{eq:mfcevolutionwithdec}) we obtained the open-loop and feedback evolutions in the described model, without and with the decoherence term, respectively. Figure \ref{fig:feedbackproc}. shows the control setup.

Now we have to define a control objective. This can be, for example, state preparation. Suppose that the initial state of the system is $\rho_0=\sum_{i=1}^{n} p_i \rho_i$. The objective is to prepare a desired eigenstate $\ket{\psi_f}$ of $F_z$ with high fidelity.

Consider the case without decoherence. We can use a simple argument to show that measurement-based feedback is superior to open-loop in the sense of dealing with uncertainties of the initial state. Note that (\ref{eq:olcevolutioncoh}) does not change the von Neumann entropy defined by
$$S(\rho)=-\tr(\rho \log(\rho))$$
This entropy is zero if and only if the state is pure and is invariant under unitary evolution. This means that we cannot prepare the target state - which is pure so has zero entropy - from an arbitrary mixed state regardless how we choose $u(t)$ and possibly $\Delta$ with the OLC evolution (\ref{eq:olcevolutioncoh}). In the MFC case, however, it has been proved (in a rather involved but rigorous way with explicit construction of the control law, see \cite{stabilization}) that we can always construct $u(t)$ such that it globally stabilizes (\ref{eq:mfcevolutionwodec}) around $\rho_f$ as $t \rightarrow \infty$. Of course, if we only consider the measurement effect (i.e. there is no control, $u(t)=0$), the evolution will project the state onto one of the eigenstates of $F_z$ and projective measurements always increase the entropy.

Now consider the case with the decoherence term. It is mathematically convenient to choose a specific representation, e.g. denote $\ket{0}=(1,0)^T$ and $\ket{1}=(0,1)^T$ the two eigenvectors of $F_z$ and also we can use the Bloch representation for $\rho$ with Bloch coordinates $x,y,z$. We can expand out the OLC and MFC evolution equations (\ref{eq:olcevolutiondecoh}) and (\ref{eq:mfcevolutionwithdec}) in terms of $x_t,y_t,z_t$ where the subscript $t$ denotes "target". We choose $\ket{\psi_f}=\ket{1}$ so $z_f=-1$. The fidelity is given by $F(\rho, \ket{\psi_f})=\frac12(1-z)$.	Now, for the OLC case, one can prove that for arbitrary initial state and arbitrary admissible control law (such that the equation (\ref{eq:olcevolutiondecoh}) has a unique solution) we have
$$\limsup_{t\rightarrow \infty} z_t \geq 0$$
which means that the fidelity is upper bounded with $F_t \leq 0.5$. So we cannot always prepare the target state with high fidelity regardless our choice of $u(t)$ and $\Delta$.

In the MFC case one can show that the fidelity is bounded by the quantity
$$F_t \leq \liminf_{t \rightarrow \infty} \left(\dfrac{1+\sqrt{l^2 + l + 1}}{2(l+1)}\right) \;\;\;\;\;\; l=\dfrac{\gamma}{M \eta}$$
so there is a quite complex connection between the efficiency of the MFC model and the parameters (decoherence strength, effective interaction strength, detection efficiency). However, this cannot tell us if $\liminf_{t \rightarrow \infty} F_t \geq 0.5$ is always true or not. The simulations in \cite{measurementbasedfeedback1} suggest that it is (at least for a large class of initial states and using an appropriate controller setup) true and the authors conclude that MFC is superior to the OLC in dealing with uncertainties.

We can extend the analysis by considering a more general form of the SME \cite{reachabilityofmfc}:
\begin{equation} \label{eq:generalsmeforanalysis}
d\rho = -i[H_0 + u(t)H_b, \rho_t]dt + M\mathcal{D}[L]\rho_t dt + \sqrt{\eta M}\mathcal{H}[L] \rho dW_t
\end{equation}
where $L$ is an arbitrary system operator (the measurement channel), $H_0$ is the effective Hamiltonian and $H_b$ is the control channel of the system. This can be easily obtained from (\ref{eq:generalsme}) by taking only one measurement channel (and call the system operator $L$), setting the Hamiltonian $H=H_0 + u(t)H_b$ and introducing the effective interaction strength $M$ or also from (\ref{eq:mfcevolutionwodec}) by setting $H_0=0 \;\;\; H_b=F_y \;\;\; L=F_z$. The corresponding OLC evolution is clearly
\begin{equation} \label{eq:generalolcforanalysis}
\frac{d \rho_t}{dt} = -i[H_0 + u(t)H_b, \rho_t] + M\mathcal{D}[L]\rho_t
\end{equation}
We assume that $H_0$ and $L$ are non-degenerate. We have to cases depending on the relationship between the system operator $L$ and the Hamiltonian $H_0$, namely if they commute or not.
\begin{theo}\label{th:olclimit}
In the OLC case with $[H_0, L]=0$ it is not possible to prepare any desired eigenstate of $H_0$ from any mixed initial state regardless how we choose $H_b,u(t)$.
\end{theo}
This can be proved in a straighforward fashion by proving that $\frac{d \tr(\rho_t^2)}{dt} \leq 0$ so we cannot increase the purity of the state with (\ref{eq:generalolcforanalysis}).
\begin{theo}
In the OLC case with $[H_0, L] \neq 0$ then $\exists$ an eigenstate $\ket{\psi}$ of $H_0$ for which the fidelity with $\rho_t$ is upper bounded, so
$$\limsup_{t \rightarrow \infty} F(\rho_t, \ket{\psi}) \leq \delta < 1$$
\end{theo}
and it is possible to construct the upper bound $\delta$ which is a function of $\ket{\psi}$ and $L$ \cite{reachabilityofmfc}.

For the MFC case let us define asymptotic reachability.
\begin{defin}
An eigenstate $\ket{\psi}$ of $H_0$ is asymptotic reachable under $L$ if $\exists \;H_b, u(t)$ such that for an arbitrary initial state $\rho_0$ there exists a unique solution of (\ref{eq:generalsmeforanalysis}), denoted by $\sigma$, which converges to $\psi$ in probability, formally:
$$\mathbb{P}\left[\lim_{t \rightarrow \infty} F(\sigma_t,\proj{\psi}{\psi})=1 \right]=1$$
\end{defin}
We can state now the following theorem.
\begin{theo}
Consider (\ref{eq:generalsmeforanalysis}) with $H_0, L$ being Hermitian and $\eta \in (0,1)$ (imperfect detection efficiency). Then every eigenstate of $H_0$ is asymptotic reachable under $L$ if and only if $[H_0, L]=0$.
\end{theo}
The proof of this is rather technical and can be found in \cite{reachabilityofmfc} (the proof is also based on the results in \cite{feedbackforstateprepmat}). We can conclude that if we choose the measurement channel $L$ appropriately, we can reach every eigenstate of $H_0$ asymptotically with the MFC model, while from Theorem \ref{th:olclimit}. we saw that in the OLC model it is not the case. In this sense, the measurement-based feedback control is superior to open-loop control in the generalized model as well.

\subsection{Quantum feedback and chaos}\label{sec:quantumchaos}
\begin{quotation}
"Anyone who uses words the "quantum" and "chaos" in the same sentence should be hung by his thumbs on a tree in the park behind the Niels Bohr Institute."
\begin{flushright}
Joseph Ford
\end{flushright}
\end{quotation}
The unravelling of the connection between chaos and quantum mechanics - despite the great interest it had gained - has been proven to be a puzzling question and was a subject to heated debates among physicists and mathematicians. As a result, there is an extensive literature on it (for a good summary, see e.g. \cite{chaosbook}) and the review of it would greatly exceed the framework if this paper and would not be relevant. Rather, we focus on the recent development of the topic which shows that Continuous measurements and feedback plays an important role in understanding the subject.

The first studies related to classical chaos date back as long as the end of the 19th century (Poincar\'e, 1892) and chaotic systems can be characterized by their exponential sensitivity to the initial conditions. This sensitivity can be measured by the Lyapunov exponent which yields the asymptotic rate of exponential divergence of two trajectories which start from neighbouring points in the phase space. If the (maximal) Lyapunov exponent is positive for a system, it is said to be chaotic. In closed quantum systems, however, the time evolution is unitary which does not allow the exponential divergence of the trajectories (in Hilbert space). In other words, the evolution of a closed quantum system is necessarily quasiperiodic (this can be directly shown from the quantum Liouville equation, for example). Quantum chaos traditionally meant the study of quantized versions of classical chaotic systems.

The paradox is apparent: how can classical mechanics emerge from quantum mechanics in an appropriate macroscopic limit if the former manifestly exhibits the above mentioned property but the latter does not? One - and the dominant - way to resolve this is the observation of the fact that every experimental setup involves measurement, therefore open quantum systems \cite{quantumchaos2}.

It was first noted in \cite{slotine} by Lloyd and Slotine that the nonlinear dynamics induced by weak quantum feedback could be used to create a novel form of quantum chaos. In this context, weak quantum feedback means that one performs a collective measurement on a large number of identical systems, thus obtains the average value of an observable while only slightly disturbing the individual systems, and then feed back this information. This form of weak measurement can be realized in NMR for example, where it is possible to monitor the induction field produced by a large number of precessing spins which gives the average value of their magnetization along a given axis but only slightly disturbs the spins. In a general picture, it formally means the following. Suppose we have $N$ identical, noninteracting quantum systems, each characterized by $\rho$. Using for example the POVM described in Section \ref{subsec:measurements}., it is possible to perform a measurement on $\rho^{\otimes N}$ which determines the single-system reduced density matrix $\rho$ to some degree of accuracy $\delta$ while disturbing it by $\epsilon$ with the property that if $N \rightarrow \infty$ then $\delta, \epsilon \rightarrow 0$. Now if we feed back this information (i.e. apply to each system a unitary transformation $U(\rho)$) then the single-system density operator will be governed by the equation $\rho'=U(\rho)\rho U^{\dagger}(\rho)$ where it is crucial to note that $U(\rho)$ can be any (possibly nonlinear) function of $\rho$. Taking the Continuous limit this immediately leads to
$$\dfrac{\partial \rho}{\partial t}=-i[H(\rho), \rho]$$
where $H(\rho)$ is the Hamiltonian corresponding to $U(\rho)$. The possible forms of $U(\rho)$ and nonlinear quantum transformations in general were analized in \cite{nonlinearquantumop}. Besides other applications, such as, for example, creating Schr\"dinger's cats, i.e. quantum systems that exist in superpositions of two quasiclassical states, systems that obey nonlinear equations could be used to create \itshape true \upshape quantum chaos; this is because the SSE need not preserve the distances of the trajectories.

A further analysis on how chaos can emerge from the SME even far from the classical limit was done in \cite{habibetal} by Habib et al. They choose the observable to be the the position operator $x$, thus the SME takes the form of (\ref{eq:smem}) with $X=x$ and consider the Duffing oscillator (single particle in a double-well potential, with sinusoidal driving) which has a Hamiltonian $H=p^2/2m + Bx^4 - Ax^2 + \Lambda x \cos(\omega t)$ where $p$ is the momentum operator, $A,B,\Lambda$ are parameters which determine the potential and the strength of the driving force. Let us label the possible realizations of the noise process $dW $ by $s$. Introduce the divergence between a fiducial trajectory and another ("shadow") trajectory infinitesimally close to it: $\Delta(t) \equiv |\mean{x(t)_{fid}} - \mean{x(t)}|$. We can now define the observatinally relevant Lyapunov exponent by
\begin{equation} \label{eq:lyapunov}
\lambda \equiv \lim_{t \rightarrow \infty} \lim_{\Delta_s(0) \rightarrow 0} \left(\dfrac{\ln(\Delta_s (t))}{t}\right) \equiv \lim_{t \rightarrow \infty} \lambda_s(t)
\end{equation}
which is reasonable because we are interested in the sensitivity of the system to changes in the initial conditions and not in the changes in the noise realizations so we keep that fixed. By simulations using paralel supercomputers they find that, after a $1/t$ behaviour, $\lambda_s(t)$ converges to a positive, finite value and this value is greater if the measurement strength $k$ is greater. From these we can conclude that there exists a purely quantum regime which evolves chaotically with a positive, finite Lyapunov exponent.

In Section \ref{sec:projqubit}. another kind of chaos is introduced which emerges from the conditional dynamics of qubits using a specific, selective protocol which can be also useful to perform control tasks. We see chaotic behaviour directly in the Hilbert-space and also in the entanglement in the multiqubit case. At the end of that section one can find a summary of feedback-induced chaos.

\subsection{Control tasks for quantum feedback control}\label{sec:controltasks}
In the following, we review some tasks which can be impleneted by QFC and - where relevant - compare their efficiency to other schemes.
\subsubsection{Stabilizing a qubit against noise using discrete time feedback and comparison with classically motivated schemes}\label{sec:stabqubitdiscrete}
Consider the following task: we prepare a qubit in one of two non-orthogonal states $\ket{\psi_1}$ and $\ket{\psi_2}$ with overlap $\scal{\psi_1}{\psi_2}=\cos \theta \;\;\; 0 \leq \theta \leq \pi/2$, for example
$$\ket{\psi_1}=\cos (\theta/2)\ket{+} + \sin (\theta/2)\ket{-} \;\;\; \ket{\psi_2}=\cos (\theta/2)\ket{+} - \sin (\theta/2)\ket{-}$$
where $\ket{\pm}=(\ket{0} \pm \ket{1})/\sqrt{2}$. In the Bloch representation \cite{quantumcomp} this means that the two states lie in the $x-z$ plane rotated by $\theta$ about the z-axis. Consider the following noise:
\begin{equation} \label{eq:noise}
\mathcal{E}_p[\rho]=p(\sigma_z \rho \sigma_z) + (1-p)\rho
\end{equation}
where $\sigma_z=\proj{0}{0} - \proj{1}{1}$ is the well-known Pauli operator and $0 \leq p \leq 0.5$. This is called a dephasing noise \cite{quantumcomp}: with probability $p$ it applies the phase flip $\sigma_z$ and with probability $(1-p)$ it leaves the system unaltered. The dephasing noise has an effect of decreasing the $x$-component of the Bloch vector. The task of stabilizing a qubit against this noise was considered in \cite{rapidprepqubit} and has recently been investigated experimentally in \cite{experimentalfeedbackcontrol} with a photonic polarization qubit. We want to find the quantum operation $\mathcal{C}$ which corrects the state after the noise has been applied and maximizes the average fidelity between the input state and the corrected state, i.e.
\begin{equation} \label{eq:max}
\max_{\mathcal{C}} \left[\dfrac{1}{2} \sum_{i=1}^2 \sand{\psi_i}{\mathcal{C}[\mathcal{E}_p[\proj{\psi_i}{\psi_i}]]}{\psi_i}\right]
\end{equation}
where $\mathcal{C}$ has to be a CPTP (completely positive and trace-preserving) map.

\begin{figure}[h!]
\begin{center}
\includegraphics[width=17cm]{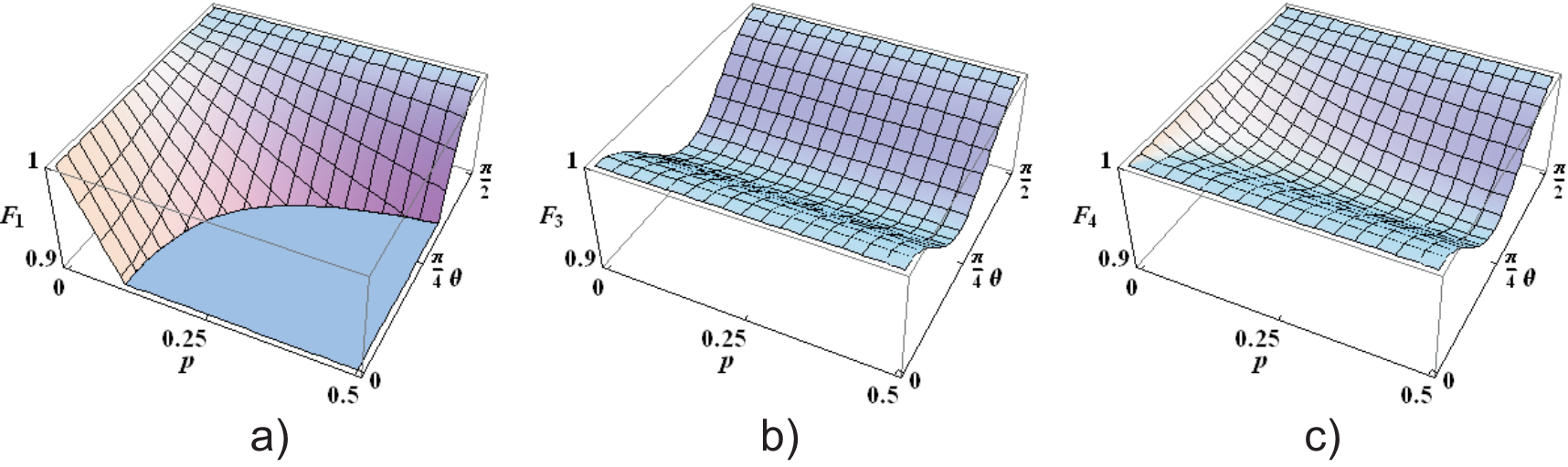}
\caption{Visualization of the fidelity functions (\ref{eq:donothing}) ("do nothing" scheme), (\ref{eq:discprep2}) ("discriminate and prepare" scheme) and (\ref{eq:quantumperformance}) ("quantum feedback" scheme).}
\label{fig:compare1}
\end{center}
\end{figure}

Now consider the first strategy: \itshape "do nothing" \upshape. We will see that in some cases this trivial strategy can be quite efficient. If we calculate the sum in (\ref{eq:max}) with $\mathcal{C}$ being the identity, the average fidelity we get is
\begin{equation} \label{eq:donothing}
F_1 = 1 - p \cos^2 \theta
\end{equation}
which is plotted on Figure \ref{fig:compare1}.a.

Now consider another - based on a classical concept - strategy: "discriminate and prepare". This means that we try to distinguish the outcoming state with a projective measurement and then prepare the state based on this result. This is a classical concept because we try to gain as much information from system as we can. The optimal projective measurement (in terms of the average probability of success) we can do succeeds with $P=\frac{1}{2}(1+\sin \theta)$ (Helstrom's measurement), independent of the noise strength $p$. Now we have to choose our states which will be prepared after the measurement. If we say that we prepare $\ket{\psi_{1,2}}$ if the measurement result is $\{1,2\}$, this yields an average fidelity of
\begin{equation} \label{eq:discprep1}
F_2 = 1 - \dfrac{1}{2}(\sin ^2 \theta - \sin ^3 \theta)
\end{equation}
However, it can be shown \cite{rapidprepqubit} that we can obtain a better average fidelity if we prepare the states
$$\ket{\phi_{\pm}}=\sqrt{\dfrac{1}{2} \pm \dfrac{\sin ^2 \theta}{2\sqrt{\sin ^4 \theta+ \cos^2 \theta}}}\ket{0} + \sqrt{\dfrac{1}{2} \mp \dfrac{\sin ^2 \theta}{2\sqrt{\sin ^4 \theta+ \cos^2 \theta}}}\ket{1}$$
i.e. we prepare $\ket{\phi_{+}}$ if we measured $\ket{\psi_1}$ and we prepare $\ket{\phi_{-}}$ if we measured $\ket{\psi_2}$. This gives an average fidelity of
\begin{equation} \label{eq:discprep2}
F_3 = \dfrac{1}{2} + \dfrac{1}{2}\sqrt{\sin ^4 \theta+ \cos^2 \theta} \geq F_2 \;\; \forall \; p,\theta
\end{equation}
which is plotted on Figure \ref{fig:compare1}.b. Its optimality can be shown using convex optimazation. In some regions (for example when $p$ is small) this scheme is outperformed by the "do nothing" scheme. Note that this is also a feedback scheme as our choice of state preparation depends on the measurement result. We say it is classical, however, because the idea is based on a classical concept, i.e. acquire as much information about the system as possible.

Now we set up a feedback control scheme with weak, non-destructive measurements. First we define our measurement operators as $E_{0,1}=M_{0,1}^{\dagger}M_{0,1}$ where
$$M_0 = \cos(\chi/2)\proj{+i}{+i}+\sin(\chi/2)\proj{-i}{-i} \;\;\; M_1 = \sin(\chi/2)\proj{+i}{+i}+\cos(\chi/2)\proj{-i}{-i}$$
where $\ket{\pm i}=(\ket{0} \pm i\ket{1})/\sqrt{2}$ (the eigenstates of $\sigma_y \equiv i\ket{1}\bra{0} - i\ket{0}\bra{1}$). This measurement can be implemented by using an ancillary qubit, a projective measurement and an entangling gate. $\chi$ is a parameter which describes the strength of the measurement: for $\chi=\pi/2$ the operator become the identity operators, for $\chi=0$ we get a projective measurement. Once we performed the measrurement, we apply an other operation based on the result (note that this is a feedback procedure):
\begin{equation} \label{eq:rot}
Z_{\eta}=e^{-i\eta \sigma_z /2}=\left( \begin{array}{cc}
e^{-i\eta  /2} & 0  \\
0 & e^{i\eta /2}  \\
\end{array} \right)
\end{equation}
We choose the angle to be $\eta$ if we obtain 0 from the measurement and $-\eta$ if we obtain 1. It can be shown that the procedure is optimal if we choose \cite{rapidprepqubit}
\begin{equation} \label{eq:eta}
\eta=\tan^{-1}[((1-2p)\cos \theta \tan \chi)^{-1}]
\end{equation}
with $0 \leq \eta \leq \pi/2$. So our correction operation altogether takes the form
\begin{equation} \label{eq:quantumcorr}
\mathcal{C}[\rho]=(Z_{+\eta} M_0)\rho (Z_{+\eta} M_0)^{\dagger} + (Z_{-\eta} M_1)\rho (Z_{-\eta} M_1)^{\dagger}
\end{equation}
and the average fidelity we get is
\begin{equation} \label{eq:quantumperformance}
F_4=\dfrac{1}{2}\left(1+ \sqrt{\cos^2 \theta + \dfrac{\sin^4 \theta}{1-(1-2p)^2\cos^2 \theta}}\right)
\end{equation}
which is plotted on Figure \ref{fig:compare1}.c. The scheme desires some interpretation. First note that the dephasing noise (\ref{eq:noise}) can be viewed as a rotation of the Bloch vector of the state by $\pm \alpha$ (with equal probability and $\alpha$ being determined by $p$) around the z-axis. Our strategy is to have a measurement which determines the sign of $\alpha$; also, we want to adjust its strength so we can vary the trade-off between information gain and back-action effect. Then we apply a feedback: we choose this to be a unitary operation which rotates it back to the desired axis based on the measurement result.

\begin{figure}[h!]
  \centering
              
  \subfloat[3D plot]{\label{fig:compare23d}\includegraphics[width=0.5\textwidth]{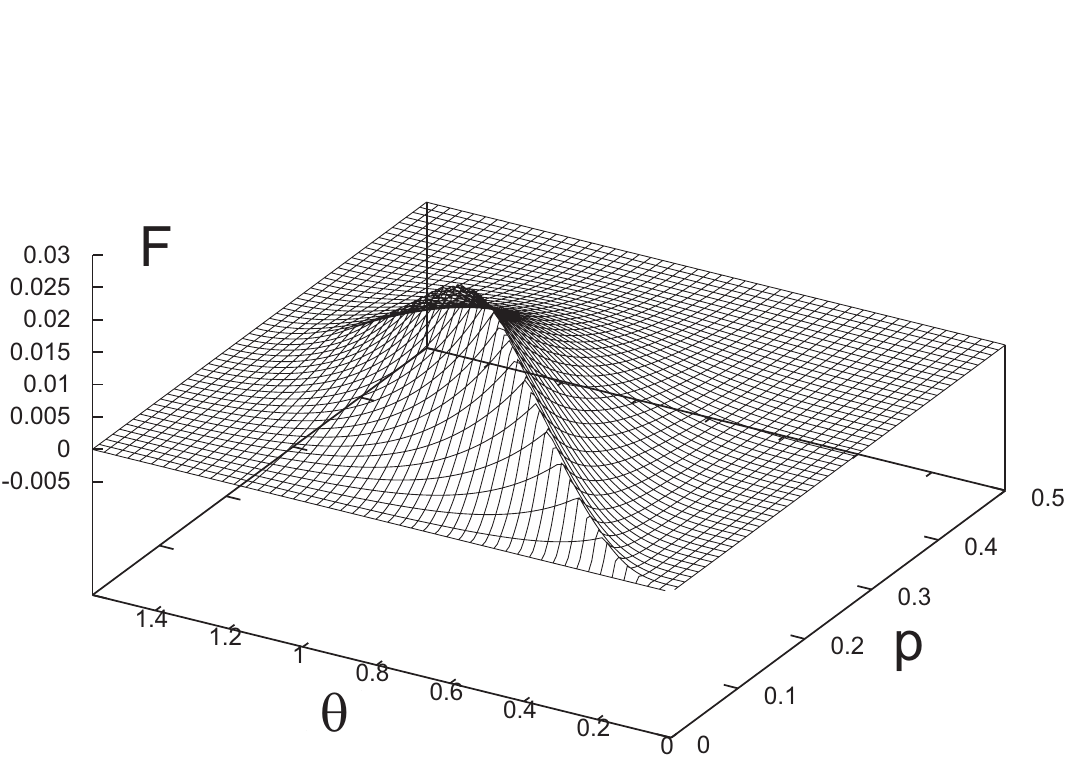}}
  \subfloat[Contour plot]{\label{fig:compare2cont}\includegraphics[width=0.5\textwidth]{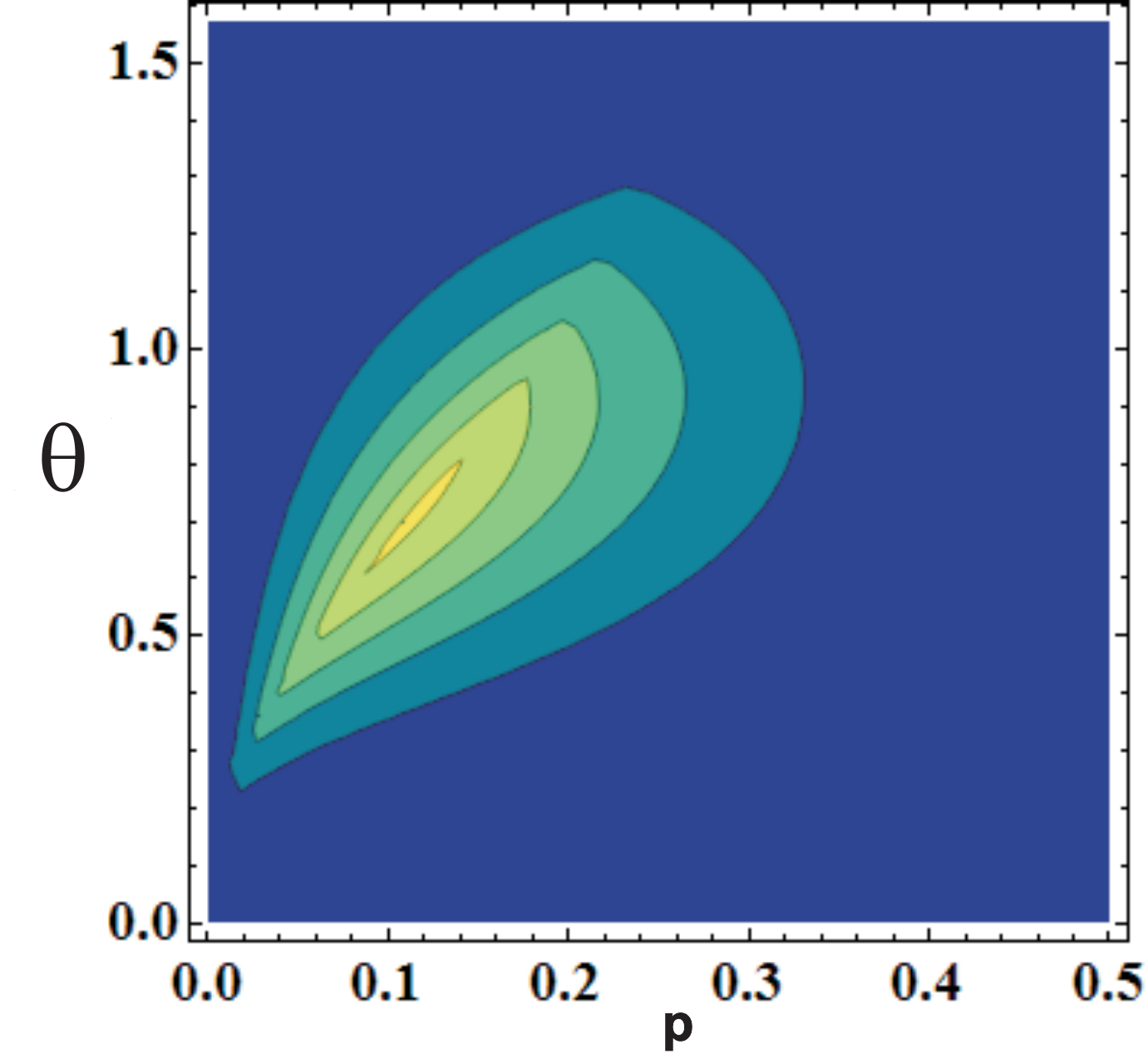}}
  \caption{Visualization of (\ref{eq:fidelcompare}). The non-destructive measurement-based feedback scheme outperforms the other schemes in the achieved average fidelity around $0.05 \lesssim p \lesssim 0.3 \;\;\; 0.3 \lesssim \theta \lesssim 1$ and it never underperforms them. The best result is at $p \approx 0.115 \;\;\; \theta \approx 0.715$ with $F=0.026$.}
  \label{fig:compare2}
\end{figure}

Figure \ref{fig:compare2}. shows the difference between the quantum feedback scheme and the other - motivated by classical control - schemes in the average fidelities, i.e.
\begin{equation} \label{eq:fidelcompare}
F(p, \theta)=F_4(p, \theta) - \max(F_1(p, \theta), F_3(p, \theta))
\end{equation}
It is apparent that $F(p, \theta) \geq 0 \; \forall \; p,\theta$ so the quantum feedback scheme always outperforms all the other schemes. It can also be shown - using the same technique as in the previous case - that for this task our feedback procedure is optimal.

\subsubsection{Rapid state purification with Continuous weak measurements}\label{sec:puriffeedback}
The following problem is very similar to the previous: the purification of a qubit system in the fastest time. However, we can make use of what we have set up in Section \ref{sec:formalism}. and perform Continuous measurements to speed up the rate of purification. There are several papers which are concerned about feedback control of two-state quantum systems \cite{twostatefeedback1,twostatefeedback2}, specifically this in \cite{rapidprepqubit}. The Continuous measurement will be performed on the $z$-component of the spin-1/2 particle (so $\sigma_z$ represents the observable) and we will use the Bloch representation with Bloch vector $\vec{a}=(a_x, a_y, a_z)^T$. With these the SME (\ref{eq:stochasticschrodingerdensity}) in terms of the Bloch components becomes
\begin{equation} \label{eq:blochevo}
da_x=-(4kdt + a_z\sqrt{8k}dW)a_x \;\;\; da_y=-(4kdt + a_z\sqrt{8k}dW)a_y \;\;\; da_z=(1-a_z^2)\sqrt{8k}dW
\end{equation}
from where we can see that - not surprisingly, as the measurement marks the $z$-direction - the relation between $a_x$ and $a_y$ is a constant of motion (the initial angle in the $x-y$ plane, $\Phi=\arctan(a_x/a_y)$ is constant). Defining $\Delta=\sqrt{a_x^2 + a_y^2}$ we can reduce (\ref{eq:blochevo}) to
\begin{equation} \label{eq:blochevodelta}
d\Delta=-(4kdt + a_z\sqrt{8k}dW)\Delta \;\;\;\;\;\; da_z=(1-a_z^2)\sqrt{8k}dW
\end{equation}
We define the impurity of the system as $\bar{p}(\rho)=1-\tr(\rho^2)=\frac12(1-\Delta^2-a_z^2)$. This is in general not a good measure of mixedness. Nonetheless, it has a simple analytical form and it is equal to the von Neumann entropy in the limit of high purity. It is possible to obtain the evolution of $\bar{p}$ from the SME using the linear quantum trajectory formulation (this is an equivalent formulation of the SME  in which the equations (\ref{eq:blochevodelta}) become linear \cite{linearfilt}) and the solution is
\begin{equation} \label{eq:impurityevo}
\bar{p}(t)=\dfrac{e^{-4kt}}{\sqrt{8 \pi t}}\int_{-\infty}^{+\infty} \dfrac{e^{-x^2/2t}}{\cosh(\sqrt{8k}x)} dx
\end{equation}
which must be solved numerically. However, we can approximate this in the long time limit (noting that if $t$ is large than the integral does not depend on $t$ and using the Taylor expansion) and we have
\begin{equation} \label{eq:impurityevo}
\bar{p}(t) \propto t^{-1/2} e^{-4kt}
\end{equation}
and in the short time limit $\bar{p}(t)$ decays exponentially with rate $4k$. So if we do not consider quantum feedback, the only way to speed up the reduction of the impurity is to increase the measurement strength $k$.

The motivation to introduce a unitary operation during the measurement is the same as in Section \ref{sec:stabqubitdiscrete}. One can calculate that the evolution of the length squared of the Bloch vector:
\begin{equation} \label{eq:blochlengthevo}
d|\vec{a}|^2 = d\Delta^2 + da_z^2=(1-a_z^2)(\Delta^2 - (1+a_z^2))8kdt + a_z((1-a_z^2)-\Delta^2)\sqrt{8k}dW
\end{equation}
and it is apparent that the best increase we can achieve is when $a_z=0$ so the Bloch vector lies in the $x-y$ plane. So, if we apply another Hamiltonian (for a time period $t$) $H=\mu(\cos(\Phi)\sigma_x - \sin(\Phi)\sigma_y)$, which generates a rotation of the Bloch vector by an angle $\alpha=2\mu t$ towards or away from the $x-y$ plane while maintaining $\Phi$, we can increase our efficiency. We can note two important things. The first one is that it is a (real-time) feedback procedure as we adjust the extra Hamiltonian at each $t$ depending on the measurement result. Secondly, that to achieve the best efficiency, we must choose the angle such that it exactly cancels the stochastic evolution which kicks out the Bloch vector from the $x-y$ plane; this, however, leads to the choice of $\mu(t) dt = \sqrt{8k}\Delta^{-1} dW$ which may require an arbitrary large Hamiltonian resource.

It is possible to solve the equation of motion and we obtain a simple result
\begin{equation} \label{eq:feedbackimpureevo}
\dfrac{d \bar{p}}{dt}=De^{-8kt}
\end{equation}
from where it is easy to see if we choose a very small target impurity (so $t$ is large), the time needed to achieve the target in the first case (which we can call classical as it is based on a classical idea) and in the quantum feedback case has a ratio of
\begin{equation} \label{eq:longtimelimitperf}
\dfrac{t_{qf}}{t_{cl}} \longrightarrow \dfrac{1}{2}
\end{equation}
So it is possible to achieve a speed-up factor of 2, in the limit when the measurement time is large compared to the measurement rate, with the help of the feedback scheme. This also allows us to perform state preparation: when the desired purity is achived, we apply a unitary on the system to rotate it to the desired target state. This is possible because we know that the Bloch vector stayed unbiased with respect to the measurement basis. The speed-up factor is a theoretical upper bound and can be less if we put constraints on the Hamiltonian. This latter case is qualitatively analyzed in \cite{rapidprepqubit}.

It was proved rigorously in \cite{optimalityofqubitprep} that this is indeed an optimal feedback procedure for the task, using Bellman equations and verification theorems. Also, it has been proven in general that in the optimal feedback control regime, it is always preferable to choose the basis of the measurement not to commute with the system density matrix \cite{feedbackofnonlinearqs}. The whole procedure was extended to the two-qubit case, where one is allowed to (weakly) measure only one of the qubits (say the first) \cite{twoqubitprep}. One might expect that the best way to purify the second qubit is to apply the optimal protocol to the first one; this, however, is not true and was falsified by a counter-example.

One can naturally ask the question: how does this speed-up change if the system is arbitrary in size? This was considered in \cite{rapidstatereduction} and it was proven that for an observable with $N$ distinct, equally spaced eigenvalues the scheme can boost the rate of purification by at least a factor of $\frac23(N+1)$ (assuming again infinitely large Hamiltonian resource). This generalized problem is significantly more involved than the qubit case and it is an open question whether the feedback procedure which achieves this performance is optimal or not.

There is another important remark. In a realistic setup we always have to consider the possible sources of delays which can affect the whole feedback loop. The total effective feedback delay is the sum of delays in the loop as the reciprocal detector bandwidth, the time needed to perform the filtering and control calculations, response time of the actuator (e.g. laser) and the electronic delays between the devices. The aforementioned feedback schemes can only work well if the dynamical timescale of the system is large compared to the effective feedback delay. Despite some remarkable developments of the devices (responsive lasers, electro-optic modulators), it is still not the case; \cite{controlimperfections} analyses the protocol when imperfections in the controls are introduced. They find that delays in the feedback loop have the most effect and for systems with slow dynamics, inefficient detection causes the biggest error. This was also the motivation for a recently proposed idea \cite{openloopfiltering} where the feedback procedure is replaced by an open-loop design together with a quantum filtering. The open-loop control is applied for some time (which time period is significantly longer than the dynamical timescale of the system) and the quantum filtering can run parallel or offline (depending on the control objective). The scheme is proven to be comparable in efficiency in several tasks (rapid measurement and purification, for instance) and much less sensitive to the delays caused by the limits of technology. 

\subsubsection{Entanglement generation}\label{sec:entgen}
\begin{quotation}
"Entanglement is iron to the classical world's bronze age."
\begin{flushright}
I.L. Chuang
\end{flushright}
\end{quotation}

Quantum entanglement is a central concept in quantum mechanics and has many applications in quantum information theory and quantum computation. With the aid of entanglement, otherwise impossible tasks may be achieved, for example in quantum communication, and it is also believed to be vital to the functioning of a quantum computer. There has been a rapid development of devices that can produce entanglement which often rely on highly controlled interactions. These can be based on trapped ions (see e.g. \cite{ionentanglement} where the authors report the creation of Greenberger-Horne-Zeilinger states with up to 14 qubits) or spatial confinement of the photons with strong atom-field coupling in a cavity, for instance. 

Applying feedback control schemes to this task was found to be useful in many cases. In fact, entanglement protection or generation is one of the most attractive applications of quantum feedback. There are a number of studies which have demonstrated that a feedback controller can effectively help the distribution of entanglement in a quantum network. Mancini and Wiseman \cite{entgenwise} showed that direct feedback can be used to enhance the correlation of two coupled bosonic modes.  The optimal measurement turns out to be nonlocal homodyne measurement in this case. Yanagisawa \cite{entgen1} presented a deterministic scheme of entanglement generation at the single-photon level between spatially separated cavities using quantum non-demolition measurement and an estimation-based feedback controller. Following these advances, in \cite{entgen2} Petersen et al. described a method to avoid entanglement sudden death in a quantum network with measurement-based feedback control. Entanglement sudden death means that entanglement completely disappears in a finite time, in which case conventional techniques - e.g. entanglement distillation - cannot assist. They consider a realistic scenario (the quantum channel is in contact with the environment and the homodyne detector has a finite bandwith) with a linear Continuous-variable cavity model. The cavities are spatially separated and the interaction is simply mediated by an optical field, in contrast to \cite{entgenwise}, where the bosonic modes interact through an optical nonlinearity.

Here, we review the control of entanglement generation between two qubits using Continuous weak measurements and local feedback in more detail. This was considered in \cite{entgen3} and, as we will see, it is an application of Jacobs' protocol (a quite tricky application, we might add) described in the previous section, so its formalism fits well in the line (see also \cite{feedbackcontroltwospin} for a more general discussion and useful introduction). Note also that the two qubit itself is a fundamental element of highly entangled states: in quantum computing, all the (unitary) entangle operations on many spins can be implemented by compositions of those on the two qubit which is referred to as the universality of quantum circuits.

Consider the density matrix $\rho$ of the two qubits which evolves according to the SME given in (\ref{eq:stochasticschrodingerdensity}). Specifically, we consider the observable to be $y=\sigma_z \otimes \sigma_z$, so the system evolves according to
\begin{equation} \label{eq:smezz}
d\rho = -k[(\sigma_z \otimes \sigma_z),[(\sigma_z \otimes \sigma_z), \rho]]dt + \sqrt{2k}((\sigma_z \otimes \sigma_z) \rho + \rho (\sigma_z \otimes \sigma_z) - 2\mean{\sigma_z \otimes \sigma_z}\rho)dW
\end{equation}
We want to quantify the entanglement of the system. For this, it is useful to expand the density operator in the Pauli basis (also called as the Fano form). For two qubits it takes the form
\begin{equation} \label{eq:fanoform}
\rho=\frac{1}{4}\sum_{i,j=\{I, X, Y, Z\}} {r_{ij} \sigma_{i} \otimes \sigma_{j}}
\end{equation}
and the coefficients $r_{ij}$ can be found as $r_{ij}=\tr(\sigma_i \otimes \sigma_j \rho)$. From the normalization condition we also know that $r_{II}=1$. It is possible to quantify the entanglement between the qubits using $R^2=\sum_{i,j} r^2_{ij}$. A pure Bell-state (maximally entangled state) has $R^2=3$; if $R^2=0$ then there is no classical correlation of the two qubits. $R^2 \leq 1$ for a product state and for a mixed state, increasing $R^2$ leads to an increase in both purity and entanglement. $R^2$ is also invariant under single qubit rotations. These equations provide the basis if one wishes to perform numerical simulations on this problem; in the following, however, we only focus on the main ideas rather than technical calculations.

Let us introduce the concept of decoherence-free subspace (DFS). A DFS is a subspace of the Hilbert space of the system that is invariant to non-unitary dynamics, i.e. it remains unaffected by the interaction of the system and its environment. This was first introduced in a quantum information theory context as these subspaces prevent destructive environmental interactions by isolating quantum information. What are the conditions for the DFS to exist? There are several possible formulations in which we can answer this question, e.g. in the Hamiltonian formulation, operator-sum representation formulation or the semigroup formulation. Here we give a description using tha latter one (as all the equations were already set up in the paper). Consider the Lindblad form of the Markovian master equation given in (\ref{eq:generalgenerator}). The dissipative part determines whether the dynamics of a quantum system will be unitary or not; in particular, when $\mathcal{D}[\rho]=0$, the dynamics will be decoherence-free. Let $\{\ket{j}\}_{j=1}^{N}$ span $\hat{\mathcal{H}}_S \subset \mathcal{H}_S$ where $\mathcal{H}_S$ is the Hilbert space of the system. Under the assumptions that the parameters $\gamma_k$ are not fine tuned and there is no dependence on the initial conditions of the initial state of the system, a necessary and sufficient condition $\hat{\mathcal{H}}_S$ to be a DFS is that all basis states $\ket{j}$ are degenerate eigenstates of the error generators (Lindbald operators). In our case we have two decoherence-free subspaces, given by $D_+ = \spann\{\ket{00}, \ket{11} \}$ and $D_- = \spann{\{\ket{01}, \ket{10} \}}$ which can be found by observing that the measurement operator $y= \sigma_z \otimes \sigma_z$ has two degenerate eigenvalues $\pm 1$. One can check that once $\rho$ is restricted to the DFS then $d \rho = 0$ according to the SME (\ref{eq:smezz}), so the measurement does not extract any useful information. Note, however, that it is easy to rotate the system out of the DFS by applying Hadamard gates locally to the qubits and this is an invertible operation. This means that it is possible to turn on and off the entanglement production procedure without turning on and off the measurement device which has practical advantages.

Now comes the essential idea. First we want to drive the system to the DFS, in which case the system will be in a classically correlated state. Once in the DFS, the system is driven towards the maximally entangled Bell state. This can be done by using only local unitary operations and the measurement of $\sigma_z \otimes \sigma_z$, however, our goal is to make use of Jacobs' protocol. Let us introduce two encoded qubits: the first will represent the extent to which information is found within the two DFS. If it is in the $\ket{0}, \ket{1}$ state then the system is confined to $D_-, D_+$, respectively. The second qubit contains the information encoded within the DFS. Physical operations can be split into two categories: the ones which commute with $\sigma_z \otimes \sigma_z$ and the ones which do not. The former operations will only affect the second encoded qubit (as these operations leave the system inside the DFS) and the latter will only affect the first encoded qubit. The basic idea is that we apply Jacobs' protocol to the encoded qubits. If we rapidly purify the first one, it means that we rapidly forced the system into the DFS. Then we apply the same protocol to the second encoded qubit. This procedure can be implemented and purifying the second encoded qubit along a specific axis generates entanglement in the physical system \cite{entgen3}. Note that the fastest rate of purification does not necessarily provide the fastest rate of entanglement generation.

The protocol described in the next section can also be used to provide entanglement generation.

\subsubsection{Projective measurement-based feedback on qubit systems and the emergence of complex chaos}\label{sec:projqubit}

In the context of quantum computation and quantum information, conditional dynamics of qubit systems have gained considerable amount of attantion, as we saw in Section \ref{sec:stabqubitdiscrete}. and \ref{sec:puriffeedback}. Here we introduce another quite exotic-looking transformation \cite{negyzetreemeles}: $\mathcal{S}: \; \rho_{ij} \longrightarrow N \rho_{ij}^2$ with the normalization factor $N=1/\sum_i \rho_{ii}^2$, so $\mathcal{S}$ simply squares the matrix elements. This transformation can be realized using basic steps involving feedback; we will restrict ourselves to qubits here. Assume we have two identical copies of the same state $\rho^{in}$ and consider the spins (qubits) pairwise: $\rho^{in} \longrightarrow \rho^{in} \otimes \rho^{in}$. Now apply the well-known XOR-gate to the pair which is
$$XOR_{12}\ket{i}_1 \ket{j}_2 = \ket{i}_1 \ket{i \oplus j}_2 \;\;\;\; i \in \{0,1\}$$
where $\oplus$ means addition mod 2. The third and last step is easy again, however, this is the key to the nonlinearity: measure the spin of the second qubit along the $z$-axis and keep the pair only if the result is "down". The whole transformation can be written in a compact form:
\begin{equation}\label{eq:strafoveg}
(\textbf{1} \otimes \mathbb{P}_0 (\mathbb{U}_{XOR}(\rho^{in} \otimes \rho^{in})\mathbb{U}_{XOR}^{\dag}) \textbf{1} \otimes \mathbb{P}_0)=\rho^{out} \otimes \mathbb{P}_0
\end{equation}
where $\mathbb{P}_0=\proj{0}{0}$. It can be easily checked that $\rho^{out}_{ij}=(\rho^{in}_{ij})^2$ indeed. The fact that $\mathcal{S}$ does not preserve the trace means that, with some finite probability, the transformation can fail. The procedure can be thought as a feedback process because it consists a filtering, based on a (projective) measurement record. There are possible generalizations of $\mathcal{S}$. For example, instead of qubits, we can use arbitrary dimensional Hilbert spaces (with the generalized XOR gate introduced in \cite{gxor}).

\begin{figure}[h!]
\begin{center}
\includegraphics[width=9cm]{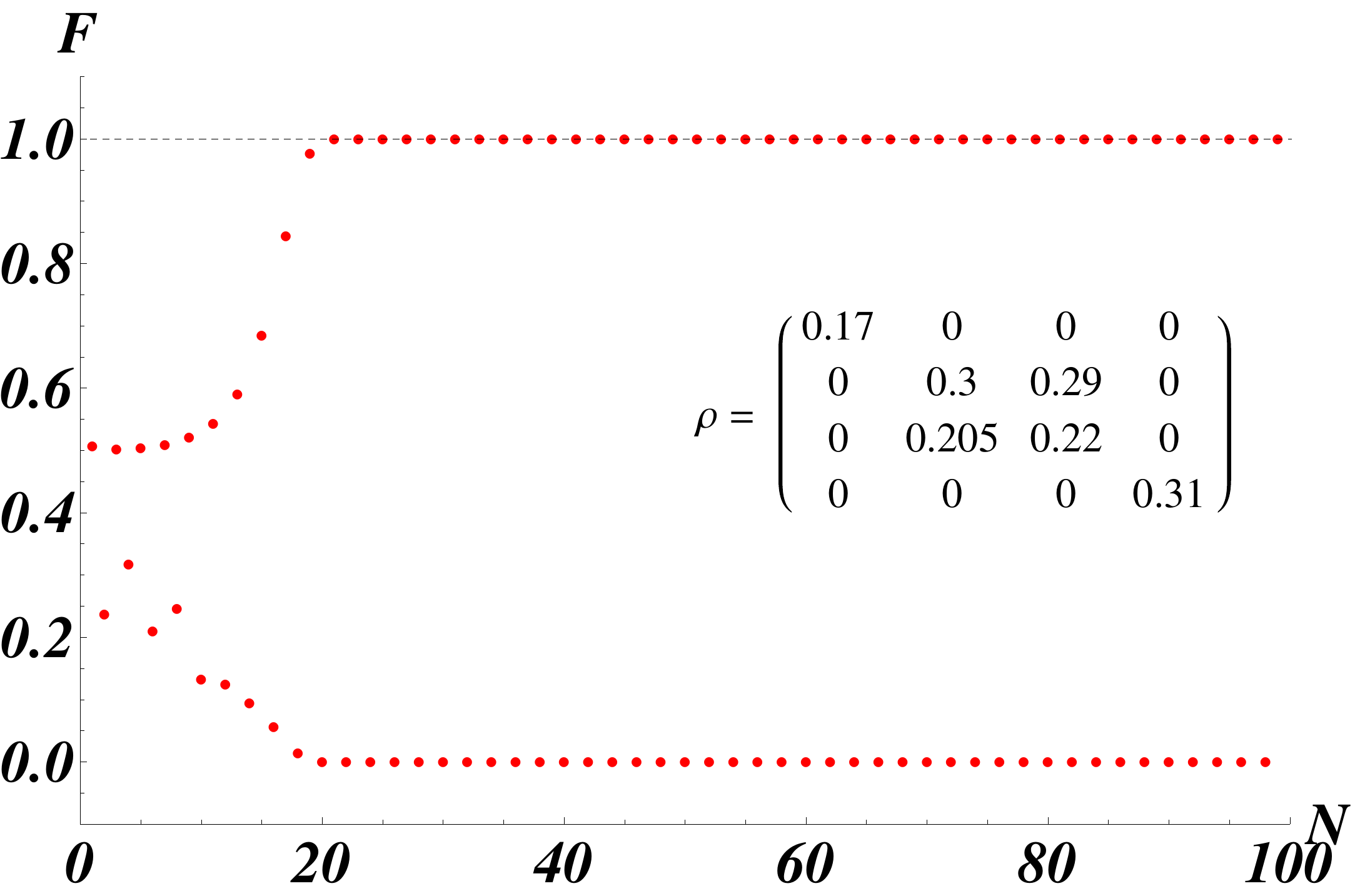}
\caption{Numerical simulation of the state purification using the map $\mathcal{F}$ iteratively. The initial fidelity is $F_0=0.5075$. After $\approx$ 20 steps, the iteration converges to the stable cycle $\{\ket{\Psi^{+}}, \ket{\Phi^{+}}\}$. Note, that the convergence is not neccesserily monotonic: after the second iteration we have $F=0.5025$.}
\label{fig:fid1}
\end{center}
\end{figure}

This strong feedback based, nonlinear transformation can be used to optimally distinguish between nonorthogonal states or to purify mixed states. As an example of the latter, let us consider a qubit pair with density matrix $\rho$. After we squared the density matrix elements with $\mathcal{S}$, we apply $U \otimes U \;\;\; U \in \mbox{SU(2)}$ (a rotation in the Hilbert space) with the parametrization
\begin{equation}\label{eq:su2}
U=\left( \begin{array}{cc}
\cos x & \sin x e^{i\varphi}  \\
-\sin x e^{-i\varphi} & \cos x  \\
\end{array} \right)
\end{equation}
and let us choose $x=\pi/4 \;\;\; \varphi=\pi/2$. One step of the whole dynamics becomes then 
\begin{equation}\label{eq:ftrafo}
\mathcal{F}[\rho]=U(\mathcal{S}\rho)U^{\dagger}
\end{equation}
 The goal is to use this transformation to restore one of the Bell states which has been perturbed (due to decoherence, for example). Define the $\ket{\Psi^{+}}$ state as
\begin{equation}\label{eq:phiplusdensity}
\ket{\Psi^{+}}= \dfrac{1}{\sqrt{2}}( \ket{10} + \ket{01})  \;\;\;\;\;\; \rho_{\ket{\Psi^{+}}}=\ket{\Psi^{+}} \bra{\Psi^{+}}=\dfrac{1}{2}\left( \begin{array}{cccc}
0 & 0 & 0 & 0 \\
0 & 1 & 1 & 0 \\
0 & 1 & 1 & 0 \\
0 & 0 & 0 & 0 \\
\end{array} \right) \end{equation}
and assume that our initial state is 
\begin{equation}\label{eq:pert1}
\rho_{pert}=\left( \begin{array}{cccc}
0.17 & 0 & 0 & 0 \\
0 & 0.3 & 0.29 & 0 \\
0 & 0.205 & 0.22 & 0 \\
0 & 0 & 0 & 0.31 \\
\end{array} \right) \end{equation}
which has a fidelity $F=Tr(\rho_{\ket{\Psi^{+}}} \cdot \rho_{pert}) = 0.5075$ with the original $\ket{\Psi^{+}}$ state. Figure \ref{fig:fid1}. is a plot of the fidelities at every iteration step. One can see that after even number of iterations the state converges to the target. This happens because $\ket{\Psi^{+}}$ is the part of the stable cycle the map. The length of this stable cycle is two, the other member being a state orthogonal to $\ket{\Psi^{+}}$. The procedure also generates entanglement in this case.

Where is the connection between this protocol and chaos? This question was raised in \cite{complexqubit1} and further analyzed in \cite{complexqubit2,complexqubit3}. Let us go back to the one qubit case (the transformation is still $\mathcal{F}$, defined in (\ref{eq:ftrafo})). Let us choose the initial state to be a pure state which we can write in the Riemann representation as
$$\ket{\psi}=N\left(z\ket{0} + \ket{1}\right) \;\;\;\; N=(1+|z|^2)^{-1/2} \;\;\;\; z \in \mathbb{C} \cup \infty \equiv \hat{\mathbb{C}} $$
where $\hat{\mathbb{C}}$ is known as the Riemann sphere. It is straightforward to show that $\mathcal{F}$ transforms a pure state into a pure state and the parameter $z$ transforms as
\begin{equation}\label{eq:ztrafo}
z \rightarrow F_p (z)=\frac{z^2 + p}{1-p^{*}z^2} \;\;\;\; p=\tan x \cdot e^{i \varphi}
\end{equation}
What we obtained is a nonlinear map on the Riemann sphere with one complex parameter $p$. These maps have been studied in detail since the beginning of the 20th century by Fatou, who studied particularly the map $z \rightarrow z^2/(z^2+2)$ \cite{fatou} and later on by G. Julia and B. Mandelbrot, just to name a few who contributed to the subject. They showed that even the simplest nonlinear maps on complex numbers can show extremely rich structures. For example, the famous Mandelbrot set emerges from the rather simple-looking map $z \rightarrow z^2 + p$.

\begin{figure}[h!]
\begin{center}
\includegraphics[width=5cm]{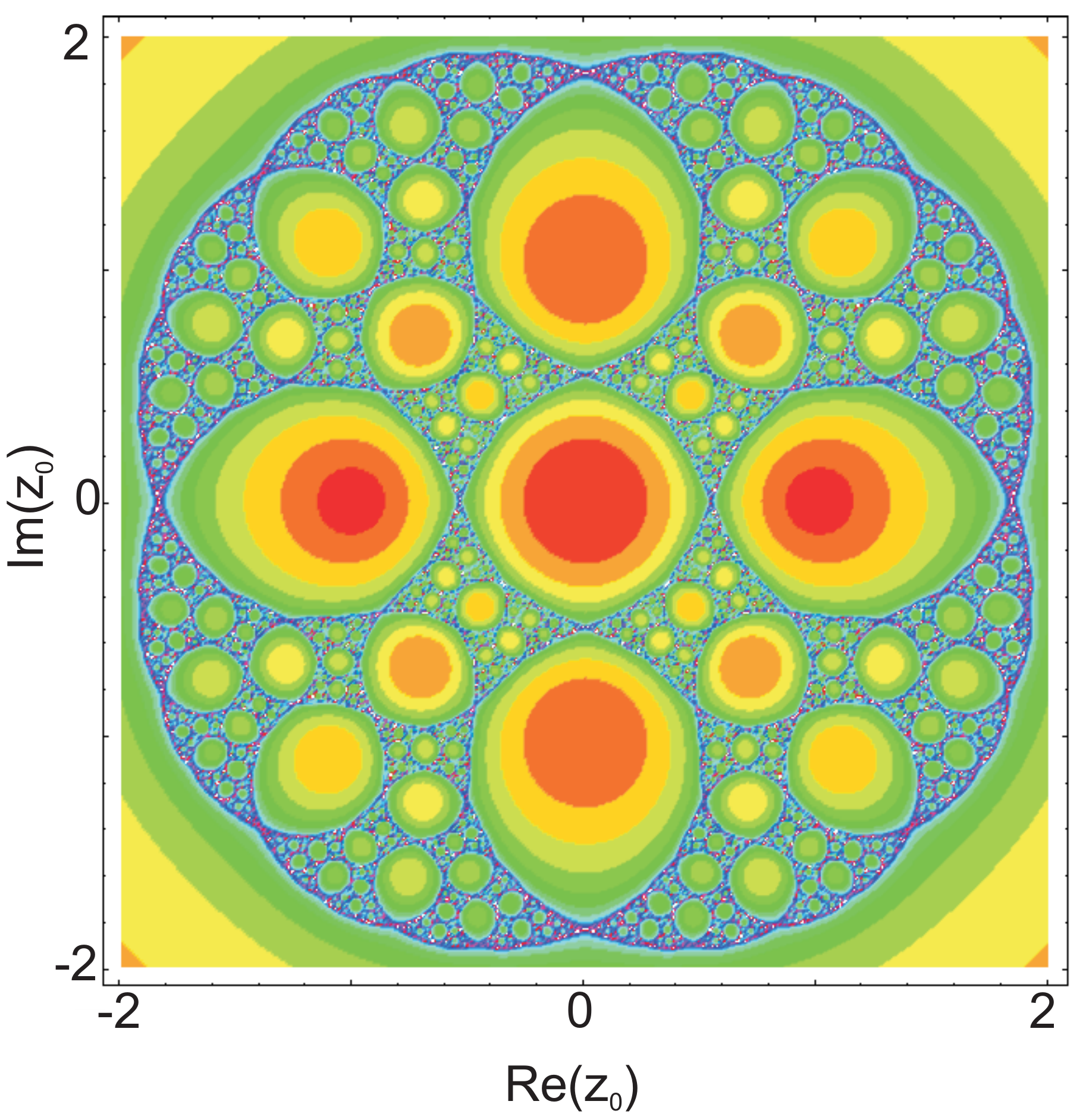}
\caption{The Julia set for $p=1$. Red means fastest, green means slowest convergence to the stable cycle (which can be proven to be the only stable cycle). In the blue domains the iteration does not converge under the criteria of the program}
\label{fig:julia1}
\end{center}
\end{figure}

$F_p(z)$ is a quadratic rational map, thus its Julia set - the set of irregular points - is non-vacous (for a rigorous treatment of dynamics of complex maps, see \cite{milnor}). This is a usual definition for complex-valued maps to be considered chaotic. For example, consider the case with $p=0$ so $F_p(z)=z^2$. The Julia set is trivial in this case: the unit circle. With a definition analogous to (\ref{eq:lyapunov}) we can calculate the Lyapunov exponent and find that it is a positive value \cite{complexqubit1}. In this sense, we can conclude that our projective measurement-based feedback protocol, if applied iteratively, can lead to true chaos in the mathematical sense. Figure \ref{fig:julia1}. shows the rich structure of the Julia set, using reproduced simulations. The program iterates the map $F_p(z)$ for a given parameter $p$ and calculates the number of steps needed to reach the stable cycle for different initial states. One can observe the fractal-like structure which is a usual property of chaotic systems; it shows that the convergence properties can change on arbitrarily small scales.

One could also iterate mixed states and see that the purity follows irregular dynamics. The two qubit case is much more complicated to treat analytically, as the initial state space and the parameter space are considerably larger but in a suitable representation one find chaotic behaviour in entanglement as well \cite{sajat}. 

The following table summarizes some charateristics of the aforementioned proposals for feedback-induced chaos in quantum systems (see also Section \ref{sec:quantumchaos}.).

\begin{center}
\begin{tabular}{|c||c|c|c|c|c|}
\hline
\multicolumn{6}{c}{Comparison} \\ \hline
\rowcolor[gray]{.94} & Measurement & Classical lim. & Stochastic & Space & Feedback \\ \hline
Lloyd and Slotince \cite{slotine}  & weak & no? & ? & ? & yes \\ \hline
Habib et al. \cite{habibetal} & Continuous & yes & yes & $\mathbb{R}^n \rightarrow \mathbb{R}^n $ & no \\ \hline
Qubit dynamics \cite{complexqubit1} & projective & ? & no & $\hat{\mathbb{C}} \rightarrow \hat{\mathbb{C}}$ & yes \\ \hline
\end{tabular}
\end{center}

\newpage
\section{Conclusion and outreach}

In this paper, some paradigms for quantum feedback control were reviewed. As quantum feedback control has been a rapidly growing research area at least for two decades now and therefore is a huge field with extensive literature, the goal was not to survey it as a whole. Rather, to give a fairly self-contained description of selected tasks which can be efficiently done using quantum feedback and to treat them in a consistent formalism. Where it was relevant, comparisons to other control designs were made and we can conclude that for many problems, quantum feedback provides optimal results. However, some of these results do not take into account the delays which are inevitably present in an experimental setup. With practical considerations, it is possible that quantum feedback, at least measurement-based quantum feedback loses its superiority against conventional methods. This is part of the reason why the field of coherent feedback networks and control is coming into the focus (for a survey see \cite{quantumnetworks}). We can also conclude that quantum feedback is linked to fundamental theoretical questions and can institute novel forms of quantum chaos. In fact, in order to have a satisfactory (and practically relevant) quantum mechanical description of the system dynamics, we need the evolution of systems which are being measured. Therefore the evolution of states is naturally conditioned on measurement results in any experimental setup which gives more understanding in the quantum - classical correspondence.

\vspace{10mm}

\textbf{Acknowledgements}
The author acknowledges the support from Sophie Schirmer for the supervision, valuable discussions and suggestions and Tamas Kiss for contionous mentoring throughout the years. Without them this work could not have been completed.

\end{document}